\documentclass[preprints,article,accept,oneauthor,10pt,a4paper]{mdpi} 

\firstpage{1} 
\pubvolume{xx}
\issuenum{1}
\articlenumber{1}
\pubyear{2018}
\copyrightyear{2018}
\history{Received: date; Accepted: date; Published: date}

\def\p{\partial}
\def\a{\alpha}
\def\b{\beta}
\def\d{\delta}
\def\D{\Delta}
\def\e{\varepsilon}

\def\l{\lambda}
\def\L{\Lambda}
\def\M{{\cal M}}
\def\rr{\varrho}
\def\g{\gamma}

\def\o{\omega}

\def\z{\zeta}

\def\ra{\rightarrow}

\usepackage{amssymb}
\usepackage{bm}

\Title{The Fractal Geometry of the Cosmic Web and its Formation}

\Author{Jose Gaite\orcidA{}}

\AuthorNames{Jose Gaite}

\address[1]{%
Applied Physics Dept.,
ETSIAE, Univ.\ Polit\'ecnica de Madrid, E-28040 Madrid, Spain; jose.gaite@upm.es}

\corres{Correspondence: jose.gaite@upm.es; Tel.: +x-xxx-xxx-xxxx}

\abstract{The cosmic web structure is studied with the
concepts and methods of fractal geometry, employing 
the adhesion model of cosmological dynamics as a basic reference. 
The structures of matter clusters and cosmic voids 
in cosmological N-body simulations or the Sloan Digital Sky Survey
are elucidated by means of multifractal geometry. 
A non-lacunar multifractal geometry can encompass three fundamental descriptions of 
the cosmic structure, namely, the web structure, hierarchical clustering, and halo 
distributions. Furthermore, it explains our present knowledge of cosmic voids. 
In this way, a unified theory of the large-scale structure of the universe seems to emerge. 
The multifractal spectrum that we obtain 
significantly differs from the one of the adhesion model and 
conforms better to 
the laws of gravity. 
The formation of the cosmic web is best modeled as a type of turbulent dynamics,
generalizing the known methods of Burgers turbulence.}


\begin{document}

\section{Introduction}

The evolution of the universe at large is ruled by gravity.  
Although the Einstein equations of the theory of general relativity are nonlinear and difficult to solve, the early evolution of the universe can be described by 
an exact solution, namely, the FLRW solution, plus its {\em linear} perturbations. 
These perturbations
grow, and they grow faster on smaller scales, becoming nonlinear. Then, 
large overdensities arise that decouple from the global expansion. 
This signals the end of a dust-like description 
of the matter dynamics and the need for a finer description of it, 
in which dissipative processes, in particular, play a prominent role. 
The dissipation is essentially a
transfer of kinetic energy from one scale to another smaller scale, due to
nonlinear mode coupling,  
as is characteristic of fluid turbulence.

Indeed, the successful {\em adhesion model} of large-scale structure formation
is actually a model of turbulence in irrotational pressure-less flow \cite{Shan-Zel,GSS}. 
This model generates the 
well-known {\em cosmic web} structure. The geometry and formation of this
web structure is the subject of this study. 
The cosmic web is a foam-like structure, formed by a web of
sheets surrounding voids of multiple sizes. 
In fact, the range of sizes is so large that we can speak of a self-similar structure.
This motivates its study by means of fractal geometry.

In fact, fractal models of the universe predate the discovery of the cosmic web structure
and arose from the idea of a hierarchy of galaxy clusters that continues 
indefinitely towards the largest scales, an idea championed by Mandelbrot \cite{Mandel}.
In spite of the work of many cosmologists along several decades, e.g., 
Peebles \cite{Pee,Peebles}, 
the debate about the scale of transition to homogeneity 
is not fully settled \cite{Pee,Peebles,Cole-Pietro,Borga,Sylos-PR,I0,Jones-RMP}.
However, the magnitude of the final value of the scale of homogeneity shall not
diminish the importance of
the fractal structure on lower scales, namely, 
in the highly nonlinear clustering regime \cite{Pee-PhysD}.
At any rate, the notion of a hierarchy of disconnected matter clusters 
has turned out to be naive. Mandelbrot \cite{Mandel} already considered 
the early signs of cosmic voids and of filamentary structure in the universe 
and, with this motivation, he proposed the study of fractal {\em texture}.  
However, only the study of the cosmological dynamics of 
structure formation can unveil the precise nature of the intricate cosmic web geometry.

The geometry of the cosmic web has been observed 
in galaxy surveys \cite{EJS,ZES,Ge-Hu} and cosmological $N$-body simulations
\cite{KlyShan,WG,Kof-Pog-Sh-M}. In addition, 
the fractal geometry of the large scale structure has been also studied
in the distribution of galaxies 
\cite{Pietronero,Jones,Bal-Schaf} and in 
cosmological $N$-body simulations \cite{Valda,Colom,Yepes}.
Recent $N$-body simulations have very good resolution and they clearly 
show both the morphological and self-similar aspects of the cosmic web.
However, while there seems to be a consensus about the reality and importance
of this type of structure, a comprehensive geometrical model is still missing. 
In particular, although our present knowledge supports a multifractal model, 
its spectrum of dimensions is only partially known, as will be discussed in this paper.

We must also consider {\em halo models} \cite{CooSh}, 
as models of large scale structure that are constructed from 
statistical rather than geometrical principles.
In the basic halo model, the matter distribution  
is separated into a distribution within ``halos'', with
given density profiles, and a distribution of halo centers in space.
Halo models 
have gained popularity as models suitable for analyzing
the results of cosmological $N$-body simulations and can  
even be employed for designing simulations \cite{Mona}.
Halo models are also employed for the study of the small scale problems 
of the Lambda-cold-dark-matter (LCDM) cosmology \cite{Popo}. Remarkably, 
the basic halo model, with some modifications, can be combined with fractal models
\cite{fhalos,I4,MN,ChaCa,Bol}.  
At any rate, halo models have questionable aspects and a careful analysis of 
the mass distribution within halos shows that it is too influenced by 
discreteness effects intrinsic to $N$-body simulations
\cite{I-Galax,Bol}.

We review in this paper a relevant portion of the efforts to understand 
the cosmic structure, with a bias towards methods of fractal geometry, as regards 
the description of the structure, and towards nonlinear methods of
the theory of turbulence, as regards the formation of the structure. 
So this review does not have as broad a scope as, for example, 
Sahni and Coles' review of nonlinear gravitational clustering
\cite{Sahni-C}, in which one can find information about other topics, such as
the Press-Schechter approach, the Voronoi foam model, the BBGKY statistical hierarchy, etc.
General aspects of the cosmic web geometry  
have been reviewed by van de Weygaert \cite{Rien}.
Here, we shall initially focus on the adhesion approximation, as
a convenient model of the cosmic web geometry and of the formation of structures 
according to methods of the theory of fluid turbulence, namely, 
the theory of {\em Burgers turbulence} in pressure-less fluids \cite{Frisch-Bec,Bec-K}.
However, the adhesion model is, as we shall see, somewhat simplistic 
for the geometry as well as for the dynamics. For the former, we need 
more sophisticated fractal models, and, for the latter,
we have to consider, first, a stochastic version of the adhesion model and, eventually, 
the full nonlinear gravitational dynamics.
 
\section{The Cosmic Web Geometry}

In this section, we study the geometry of the universe on middle to large scales,
in the present epoch, and from a descriptive point of view. As already mentioned, there 
are three main paradigms in this respect: the cosmic web that arises from the 
Zeldovich approximation and the adhesion model \cite{Shan-Zel,GSS}, 
the fractal model \cite{Mandel},
and the halo model \cite{CooSh}. These paradigms have different origins and motivations and 
have often been in conflict. Notably, a fractal model with no transition to homogeneity, as Mandelbrot proposed \cite{Mandel}, is in conflict with the standard FLRW cosmology and, 
hence, with the other two paradigms. Indeed, Peebles' cosmology textbook \cite{Peebles} places 
the description of ``Fractal Universe'' in the chapter of {\em Alternative Cosmologies}. 
However, a fractal nonlinear regime in the FLRW cosmology does not generate any conflict 
with the standard cosmology theory \cite{Pee-PhysD} 
and actually constitutes an important aspect of the cosmic web geometry. Furthermore, 
we propose that the three paradigms can be unified and that they represent, 
from a mature point of view, 
just different aspects of the geometry of the universe on middle to large scales.
Here we present the three paradigms, separately but attempting to highlight
the many connections between them and suggesting how to achieve a unified picture.

\subsection{Self-similarity of the Cosmic Web}
\label{ss_web}

Although the cosmic web structure has an obviously self-similar appearance, this aspect 
of it was not initially realized and it was instead assumed that there is a 
``cellular structure'' with a limited range of cell sizes \cite{cellular}. However, 
the web structure is generated by Burgers turbulence, and in the 
study of turbulence and, in particular, of Burgers turbulence, it is natural to assume 
the self-similarity of 
velocity correlation functions \cite{Frisch-Bec,Bec-K}. Therefore, it is natural to look for 
scaling in the cosmic web structure.
The study of gravitational clustering with methods of the theory of turbulence  
is left for Sect.~\ref{form}, and
we now introduce some general notions, useful to understand the 
geometrical features of the cosmic web and to connect with the fractal and halo models.

The FLRW solution is unstable because certain perturbations grow.
This growth means, for dust (pressureless) matter following geodesics, 
that close geodesics either diverge or converge. In the latter case, the geodesics 
eventually join and matter concentrates on {\em caustic} surfaces. In fact, 
from a general relativity standpoint, caustics are 
just a geometrical phenomenon that always appears in the attempt to construct a
synchronous reference frame from a three-dimensional
Cauchy surface and a family of geodesics orthogonal to this surface \cite{LL}. 
But what are, in general, non-real singularities, owing to the 
choice of coordinates, become real density singularities in the irrotational flow of dust matter.

One can consider the study and classification of the possible types of caustics as a 
purely geometrical problem, connected with a broad class of problems in the 
theory of singularities or {\em catastrophes} \cite{Arnold}. To be specific, the 
caustic surfaces are singularities of potential flows and are called 
{\em Lagrangian singularities} \cite[ch.\ 9]{Arnold}. 
A caustic is a critical point of the {\em Lagrangian map}, which maps initial 
to final positions. After a caustic forms, the inverse Lagrangian map
becomes multivalued, that is to say, the flow has several streams (initially, just three).
Although this is a general process, 
a detailed study of the formation of 
singularities in cosmology has only been carried out in the {\em Zeldovich approximation},
which is a Newtonian and {\em quasi-linear} approximation of the dynamics 
\cite{Shan-Zel,Sahni-C}.
Recently, Hidding, Shandarin and van de Weygaert \cite{Hidding} have described the
geometry of all generic singularities formed in the Zeldovich approximation, displaying 
some useful graphics that show the patterns of folding, shell crossing and multistreaming.
Since the classification of Lagrangian singularities is universal  
and so is the resulting geometry, those patterns are not restricted to the 
Zeldovich approximation.

Unfortunately, the real situation in cosmology is more complicated: the initial condition
is not compatible with a smooth flow, so the theory of singularities of smooth maps can 
only be employed as an approximation, suitable for smoothened initial conditions. 
This approach yields some results
\cite{Sahni-C}. However, a non-smooth initial condition gives rise to 
a random distribution of caustics
of all sizes, with an extremely complex distribution of multistreaming flows, 
whose geometry is mostly unexplored. 
Besides, the Zeldovich approximation fails after shell crossing. The 
real multistreaming flow generated by Newtonian gravity has been studied 
in simplified situations; for example, 
in one dimension (which can describe the dimension transverse to caustics). 
Early $N$-body simulations showed that the thickness of an isolated multistreaming zone 
grows slowly, as if gravity makes particles stick together \cite{Shan-Zel}. 
The analytical treatment of Gurevich and Zybin \cite{GZ}, for smooth initial conditions,
proves that multistreaming gives rise to power-law mass concentrations.
One-dimensional simulations with cosmological initial conditions 
(uniform density and random velocities) \cite{Aurell_1,Miller,Joyce}
show that particles tend to concentrate in narrow multistreaming zones pervading 
the full spatial domain. Furthermore, 
both the space and phase-space distributions tend to have {\em self similar} properties.

At any rate, before any deep 
study of multistreaming was undertaken, 
the notion of ``gravitational sticking'' of matter suggested a simple modification of 
the Zeldovich approximation that suppresses multistreaming.
It is natural to replace the collisionless 
particles by small volume elements and hence to assume that, 
where they concentrate and produce an infinite density, it 
is necessary to take into account lower-scale processes. The simplest way to achieve this is 
embodied in the {\em adhesion model}, which supplements the Zeldovich approximation with a 
small viscosity, giving rise to the (three-dimensional) Burgers equation \cite{Shan-Zel}. 
One must not identify this viscosity with the ordinary gas viscosity but with an effect 
of {\em coarse graining} the collisionless Newtonian dynamics \cite{BD}.
Even a vanishing viscosity is sufficient to prevent multistreaming. 
Actually, the vanishing viscosity limit is, 
in dimensionless variables, the high Reynolds-number limit, which gives rise to 
Burgers turbulence. Although we postpone the study of cosmic turbulence to
Sect.~\ref{form}, we summarize here some pertinent results of the adhesion model.

\begin{figure}[H]
\centering
\includegraphics[width=7.4cm]{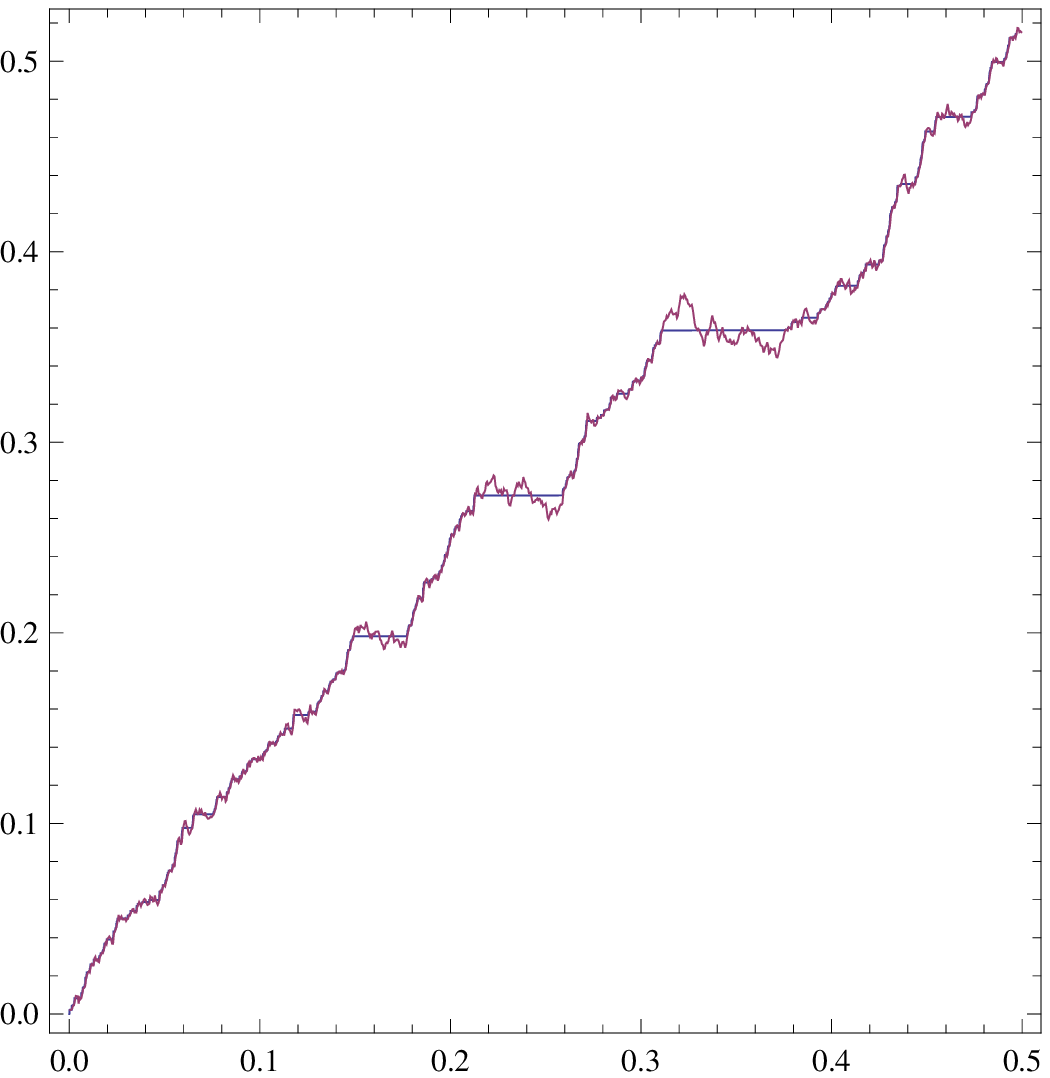}\hspace{2mm}
\includegraphics[width=7.6cm]{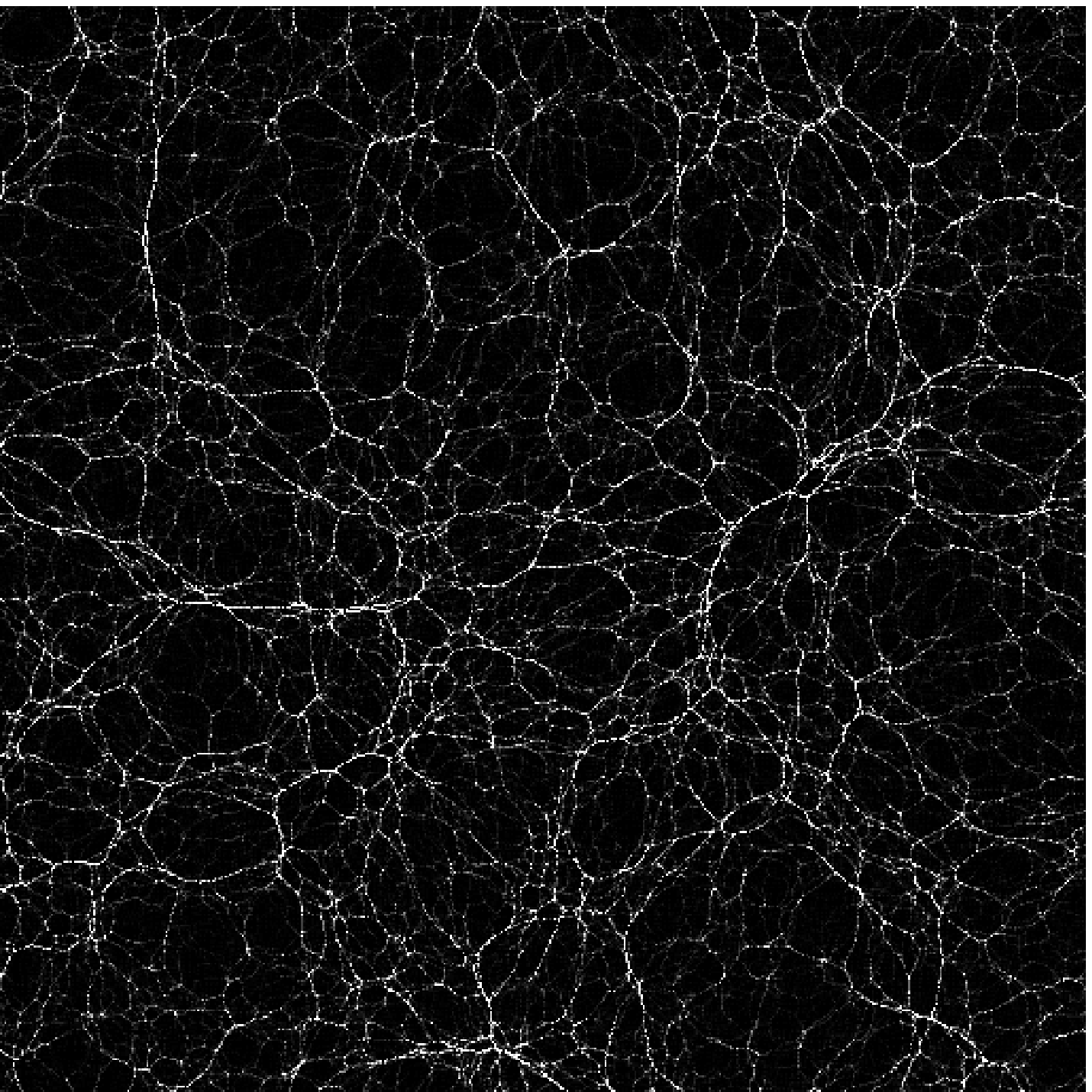}
\caption{
Realizations in one or two dimensions of the exact solution of the adhesion model 
with scale-invariant initial velocity field ($h=1/2$). 
In both, the length scale is arbitrary, due 
to the self-similarity, so that the relevant scale is the scale of the largest structures.  
Left-hand side: in one dimension, 
Lagrangian map from initial to final positions, with multistreaming,  
and devil's staircase formed by mass sticking. 
Right-hand side: web structure of final positions in two dimensions. 
}
\label{devil}
\end{figure}   

A deep study of the geometry of the mass distribution generated by the adhesion model has 
been carried out by Vergassola {\em el al} \cite{V-Frisch}. Their motivation was to compare the prediction of the adhesion model with the result of the Press-Schechter approach to 
the mass function of collapsed objects (this approach is explained in Ref.~\cite{Sahni-C}).  
But Vergassola {\em el al} actually show how
a self-similar mass distribution arises (including the {\em analytical} 
proof for a particular case). 
The adhesion model (with vanishing viscosity) has an exact geometrical solution, 
in terms of the {\em convex hull} of the {\em Lagrangian potential} \cite{V-Frisch}.
In one dimension and with cosmological initial conditions,
the adhesion process transforms 
the Lagrangian map to a random {\em devil's staircase}, with an infinite number of steps, 
whose lengths correspond to the amounts of stuck mass (Fig.~\ref{devil}, left-hand side).
To be precise, if the initial condition consists of a uniform density and a Gaussian velocity 
field with a power-law energy spectrum, then, at any time $t > 0$, singularities arise 
in the velocity field and mass condensates on them, 
giving rise to the steps of the devil's staircase.
Such condensations initially contain very small mass but grow as mass sticking proceeds, 
preserving a self-similar distribution in space and time.

Let us explain in more detail the properties of the self-similar mass distribution that
arises from a scale-invariant initial velocity field, such that 
$$\bm{u}(\l \bm{r}) = \l^h \bm{u}(\bm{r}),\; h\in (0,1),$$ 
namely, a {\em fractional Brownian} field with Hurst exponent $h$ \cite{V-Frisch}. 
These fields are Gaussian random fields that generalize the Brownian field, with $h=1/2$ and 
independent increments, in such a way that, for $h>1/2$, same-sign increments 
are likely ({\em persistence}), whereas, for $h<1/2$, opposite-sign increments 
are likely ({\em anti-persistence}) \cite[ch.~IX]{Mandel}. 
If $h \ra 1$, then the field tends to be smooth, whereas, if $h \ra 0$,
then the field tends to be discontinuous. 
A fractional Brownian velocity field corresponds to Gaussian density fluctuations 
with power-law Fourier spectrum of exponent $n=2(1-h)-d,$ in $d$ dimensions. 
There is a scale $L \propto t^{1/(1-h)}$, called ``coalescence length'' by 
Vergassola {\em el al} \cite{V-Frisch}, such that it separates small scales, where 
mass condensation has given rise to an inhomogeneous distribution, from large scales, 
where the initial conditions hold and the distribution is still practically homogeneous
($t$ is a redefined time variable, namely, the growth factor of perturbations 
\cite{Shan-Zel,V-Frisch}).
In one dimension, we have 
a devil's staircase with steps that follow a power law, 
so that the cumulative mass function is 
\begin{equation}
N(M>m) \propto m^{-h}.
\label{gap-law}
\end{equation}
However, the number of steps longer than $L$, that is to say, of large mass condensations,  
has an exponential decay (like in the Press-Schechter approach).
These results can be generalized to higher dimensions, in which the web structure is 
manifest (Fig.~\ref{devil}, right-hand side).
However, in higher dimensions, the full geometrical construction is 
very complex and hard to visualize \cite{Bernardeau-Vala}. 

Even though the results of the adhesion model are very appealing, we must notice its 
shortcomings. On the one hand, it is based on the Zeldovich approximation, which is 
so simple that the mass function is directly given by the initial energy spectrum.
The Zeldovich approximation is not arbitrary and is actually the first order of 
the Lagrangian perturbation theory of 
collisionless Newtonian dynamics, as thoroughly studied by Buchert (e.g., \cite{Buchert}); 
but some nice properties of the Zeldovich approximation are not preserved in higher orders.
At any rate, the Zeldovich approximation is exact in one dimension, yet 
the one-dimensional adhesion model is not, because the instantaneous adhesion of matter 
is not realistic and differs from what is obtained in calculations or $N$-body simulations, 
namely, 
power-law mass concentrations instead of zero-size mass condensations. 
\cite{GZ,Aurell_1,Miller,Joyce,Aurell_2}. The calculations and simulations 
also show that the final mass distribution is not simply related to the 
initial conditions.

Nevertheless, it can be asserted that 
the locations of power-law mass concentrations in cosmological $N$-body simulations, 
in one or more dimensions,  
form a self-similar distribution that looks like the cosmic web of 
the adhesion model. 
Indeed, 
all these mass distributions are actually particular types of
multifractal distributions \cite{fhalos,I4,MN,ChaCa,Miller}. 
We now study fractal models of large-scale structure from a general standpoint. 

\subsection{Fractal Geometry of Cosmic Structure}
\label{FG}

The concept of cosmic web structure 
has its origin in the adhesion model, 
which is based on the study of cosmological dynamics. In contrast, 
fractal models of large-scale structure have mostly 
observational origins, combined with ancient theoretical motivations 
(the solution of the Olbers paradox, the idea of nested universes, etc.). Actually, 
the ideas about the structure of the universe have swung between the principle of 
homogeneity and the principle of hierarchical structure \cite{Peebles}. 
The Cosmological Principle is the modern formulation of the former, 
but it admits a {\em conditional} formulation that supports the latter 
\cite[\S 22]{Mandel}. There is no real antithesis and the synthesis is provided by 
the understanding of how a fractal nonlinear regime develops below some 
homogeneity scale, consequent to the instability of the FLRW model. 
This synthesis is already present in the adhesion model, in which 
the homogeneity scale is the above-mentioned {\em coalescence length} $L$.

The notion of hierarchical clustering, namely, the idea of clusters of galaxies that 
are also clustered in 
superclusters and so onwards, inspired the early fractal models of the universe. 
This idea can be simply realized in some 
three-dimensional and random generalization of the Cantor set, with 
the adequate fractal dimension \cite{Mandel}. In mathematical terms, the model places
the focus on the geometry of {\em fractal sets}, characterized by the Hausdorff dimension.
A random fractal set model for the distribution of galaxies
has been supported by measures of the  reduced two-point
correlation function of galaxies, which is long known to be 
well approximated as the power law 
\begin{equation}
\xi(r)=\frac{\langle \rr(\bm{r})\,\rr(\bm{0})\rangle}{\langle \rr\rangle^2}-1 = 
\left({r_0\over r}\right)^{\gamma},
\label{xi}
\end{equation}
with exponent $\gamma \simeq 1.8$ and {\em correlation length}
$r_0 \simeq 6$ Mpc, as standard but perhaps questionable values 
\cite{Mandel,Pee,Peebles,Cole-Pietro,Borga,Sylos-PR,I0,Jones-RMP}. 
In fact, the name correlation length for $r_0$ is essentially a misnomer \cite{I0}, and $r_0$ 
is to be identified with the present-time value of $L$ in the adhesion model, 
that is to say, 
with the current scale of transition to homogeneity.
Although the value of $r_0$ has been very debated, 
Eq.~(\ref{xi}) implies, for any value of $r_0$, 
that there are very large fluctuations of the density 
when $r \ll r_0$, and indeed that the distribution is fractal with Hausdorff dimension 
$3-\gamma$ \cite{Mandel,Cole-Pietro,Pee,Peebles,Borga,Sylos-PR,I0,Jones-RMP,Pee-PhysD}.

It would be convenient that a fractal were described only by its Hausdorff dimension,
but it was soon evident that one sole number
cannot fully characterize the richness of complex fractal structures and, in particular, 
of the observed large-scale structure of the universe. Indeed, fractal sets
with the same Hausdorff dimension can {\em look} very different and 
should be distinguishable from a purely mathematical standpoint. This is obvious 
in two or higher dimensions, where arbitrary subsets possess {\em topological invariants} 
such as connectivity that are certainly relevant. Furthermore, it is also true 
in one dimension, in spite of the fact that any (compact) fractal subset of $\mathbb{R}$ 
is totally disconnected and has trivial topology. 

In an effort to complement the concept of fractal dimension, 
Mandelbrot was inspired by the appearance of galaxy clusters and superclusters and
introduced the notion of {\em lacunarity}, without
providing a mathematical definition of it \cite[\S 34]{Mandel}. 
Lacunarity is loosely defined as the property of having large gaps or voids ({\em lacuna}). 
This property gained importance in cosmology with the discovery of large voids in 
galaxy redshift surveys, which came to be considered a characteristic feature of 
the large-scale structure of the universe \cite{ZES,Ge-Hu}. The precise nature 
of cosmic voids is indeed important to characterize the geometry of the cosmic web and 
will be studied below.

In general, for 
sets of a given Hausdorff dimension, there should be a number of parameters that specify 
their appearance, what Mandelbrot calls {\em texture} \cite[ch.~X]{Mandel}.
Lacunarity is the first one and can be applied to sets in any dimension,
including one dimension. Others are only applicable in higher dimension; for 
example, parameters to measure the existence and extension of subsets of a fractal set 
that are topologically 
equivalent to line segments, an issue related to percolation \cite[\S 34 and 35]{Mandel}.

In fact, the methods of percolation theory have been employed in the study of the cosmic web 
\cite{Shan-Zel}. Those methods are related to the mathematical notion of path connection of 
sets, which is just one topological property among many others. 
Topological properties are certainly useful, but
topology does not discriminate enough in fractal geometry, 
as shown by the fact that all (compact) fractal subsets 
of $\mathbb{R}$ are topologically equivalent. According to Falconer \cite{Falcon}, 
fractal geometry can be defined in analogy with topology, replacing 
bi-continuous transformations by a subset of them, namely, 
{\em bi-Lipschitz transformations}. The Hausdorff dimension is invariant under 
these transformations, although it is not invariant under arbitrary 
bi-continuous transformations. Therefore, the set of parameters characterizing a 
fractal set includes, on the one hand, topological invariants and, on the other hand, 
properly fractal parameters, beginning with the Hausdorff dimension. 
Unfortunately, there has been little progress in the definition of these parameters, 
beyond Mandelbrot's heuristic definition of texture parameters.

Besides the need for a further development of the geometry of fractal sets, 
we must realize 
that the concept
of {\em fractal set} is not sufficient to deal with the complexity of cosmic geometry.
This is easily perceived 
by reconsidering the mass distribution generated by the adhesion model, in particular, 
in one dimension. In the geometry of a devil's staircase, we must distinguish
the set of lengths of the steps from the locations of these steps, which are 
their ordinates in the graph on the left-hand side of Fig.~\ref{devil}. 
These locations constitute a 
{\em dense} set, that is to say, they leave no interval empty \cite{V-Frisch}. 
The set is, nevertheless, a countable set and therefore has zero Hausdorff dimension. 
But more important than the locations of mass 
condensations is the magnitude of these masses (the lengths of the steps). 
To wit, we are not just 
dealing with the {\em set} of locations but with the full {\em mass distribution}. 
Mandelbrot \cite{Mandel} does not emphasize
the distinction between these two concepts and actually 
gives no definition of mass distribution;
but other authors do: for example, Falconer includes an introduction to
{\em measure theory} \cite{Falcon}, which is useful to precisely define 
the Hausdorff dimension as well as to define the concept of mass distribution. 
A mass distribution is a finite measure on 
a bounded set of $\mathbb{R}^n$, a definition that requires some mathematical background
\cite{Falcon}, although the intuitive notion is nonetheless useful.

The geometry of a generic mass distribution on $\mathbb{R}$ is easy to picture, 
because we can just consider the geometry of the devil's staircase and generalize it. 
It is not difficult to see that the {\em cumulative} mass 
that corresponds to a devil's staircase interval is given by switching its axes; for example, 
the total mass in the interval $[0,x]$ is given by the length of the initial interval 
that transforms to $[0,x]$ under the Lagrangian map. The switching of axes can be 
performed mentally on Fig.~\ref{devil} (left). The resulting 
cumulative mass function is monotonic but not continuous. 
A general cumulative mass function needs only to be monotonic, 
and it may or may not be continuous: 
its discontinuities represent mass condensations (of zero size).
Given the cumulative mass function, we could derive the mass 
density by differentiation, in principle. This operation should present 
no problems, because a monotonic function is almost everywhere differentiable.
However, the inverse devil's staircase has a {\em dense} set of discontinuities, 
where it is certainly non-differentiable. This does not constitute a contradiction, 
because the set of positions of discontinuities is countable and therefore has 
zero (Lebesgue) measure, 
but it shows the complexity of the situation. Moreover, in the complementary set 
(of full measure), where the inverse devil's staircase is differentiable, 
the derivative vanishes \cite{V-Frisch}.

In summary, we have found that our first model of cosmological 
mass distribution is such 
that the mass density is either zero or infinity. We might attribute this singular 
behavior to the presence of a dense set of zero-size mass condensations, due to the adhesion approximation.
Therefore, we could expect that a generic mass distribution should be non-singular.
Of course, the meaning of ``generic'' is indefinite. From a mathematical standpoint, 
a generic mass distribution 
is one selected at random according to a natural probability distribution 
(a probability distribution on the space of mass distributions!). This apparently 
abstract problem is of interest in probability theory and has been studied, with
the result that the standard methods of randomly generating  
mass distributions indeed produce {\em strictly singular} distributions
\cite{Monti}.
This type of distributions have no positive finite density anywhere, like 
our discontinuous example, but the result also includes {\em continuous} 
distributions.
In fact, the mass distributions obtained in cosmological
calculations or simulations 
\cite{GZ,Miller} seem to be continuous but strictly singular nonetheless; namely,
they seem to contain dense sets of power-law mass concentrations. 

\begin{figure}
\centering
\includegraphics[width=6.5cm]{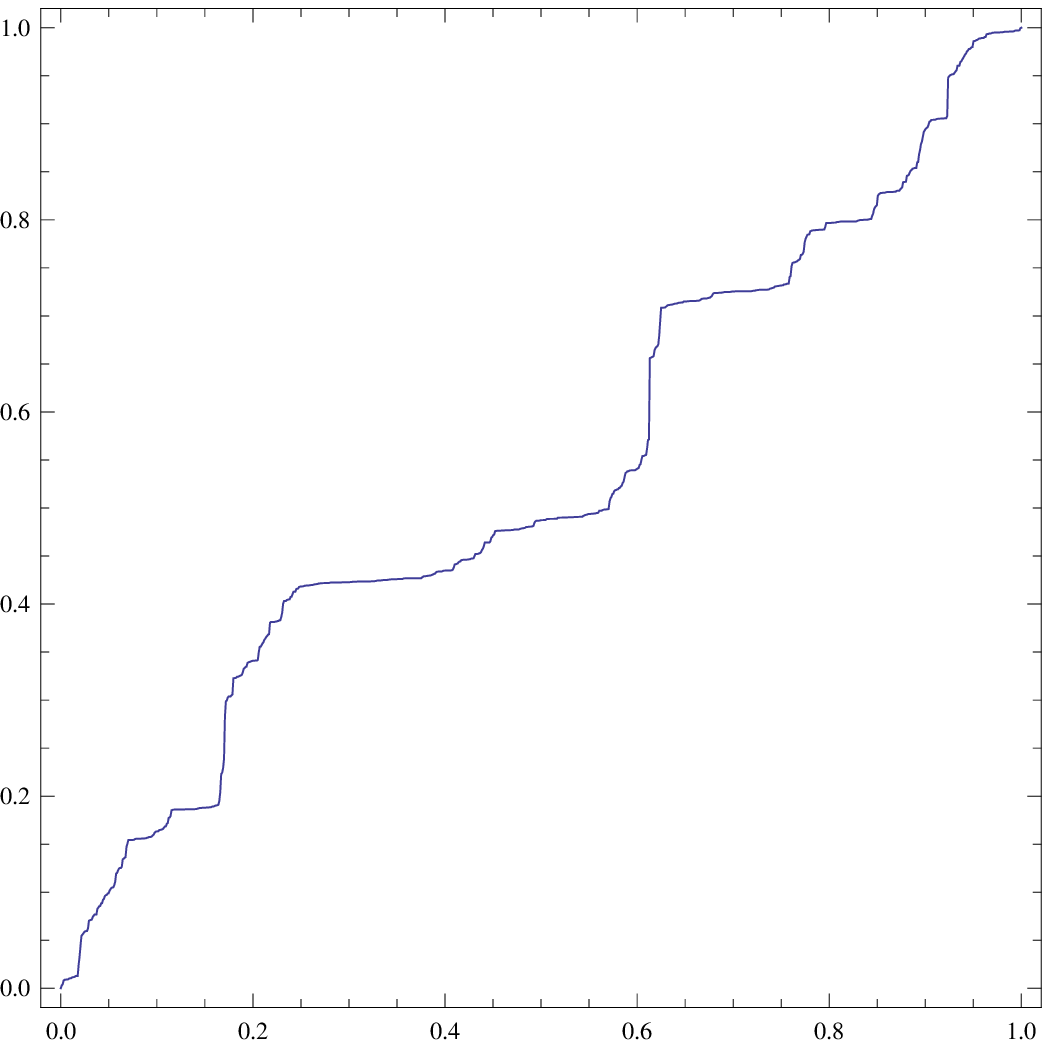}\hspace{2mm}
\includegraphics[width=8cm]{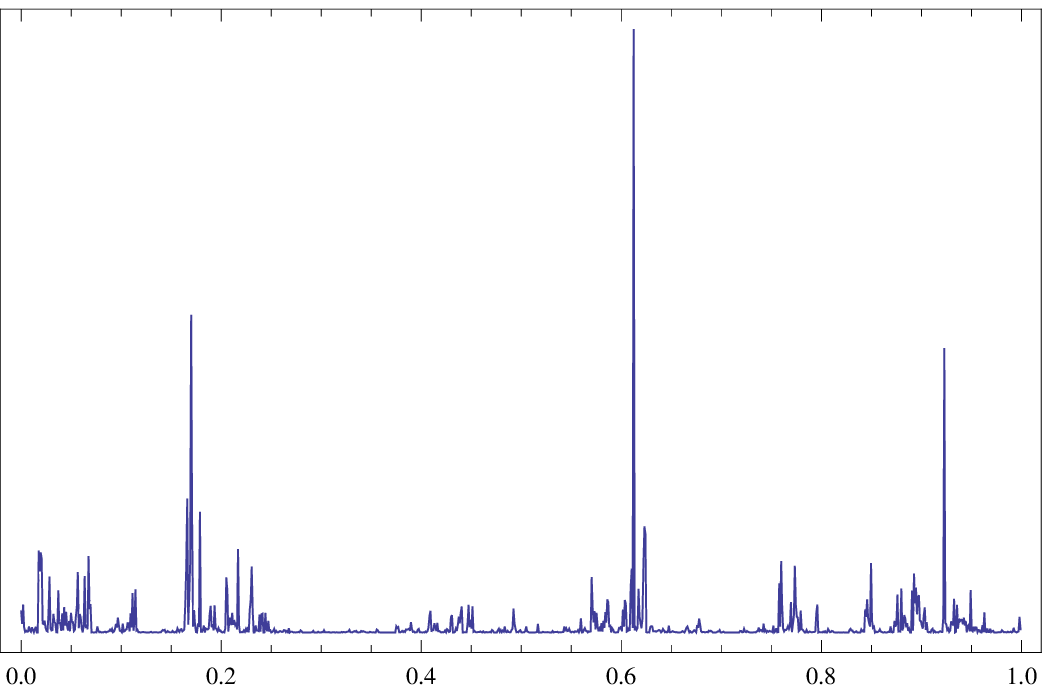}
\caption{Left: One example of random mass distribution function, 
generated as a random homeomorphism of the interval $[0,1]$. 
Right: The derivative of this function, giving the mass distribution.}
\label{G-homeo}
\end{figure}   

Let us consider the example of random mass function
displayed in Fig.~\ref{G-homeo},  
which is generated with one of the methods referred to in \cite{Monti}. 
Actually, the real mass distribution, namely, the mass density, is
obtained as the derivative of the mass function and 
is not a proper function, because it takes only the values zero or infinity. The graphs 
in Fig.~\ref{G-homeo} are the result of coarse graining with a small length, but it is 
evident that the values of the density are either very large or very small. 
Furthermore, the large or small values are clustered, respectively. They form mass 
clusters or voids, 
the former occupying vanishing total length and the latter occupying the full total length, 
in the limit of vanishing coarse-graining length. Unlike the 
inverse devil's staircase, the present mass distribution function is continuous, 
so its graph contains no vertical segments, in spite of being very steep 
and actually non-differentiable at many points. 
Therefore, there are no zero-size mass condensations but just mass concentrations. 
These mass concentrations are clustered (in our example, 
as a consequence of the generating process).
In view of its characteristics, this random mass function illustrates the type 
of one-dimensional mass distributions in cosmology, which can be considered generic. 
Some further properties of generic strictly singular mass distributions are worth 
noticing. 

To any mass distribution is associated a particular set, namely, 
its supporting set, which is the smallest {\em closed} set 
that contains all the mass \cite{Falcon}. This set is the full interval in the 
examples of Fig.~\ref{devil} or Fig.~\ref{G-homeo}; but it can be smaller in 
strictly singular mass distributions and it can actually 
be a fractal set, namely, a set with Hausdorff dimension strictly smaller than one. 
In fact, if Mandelbrot \cite{Mandel} seldom distinguishes between 
fractal sets and fractal mass distributions, it is probably because, 
in most of his examples, the support
is a fractal set and the mass is {\em uniformly} distributed on it. In this simple situation, 
the distinction is superfluous. However, we have to consider singular mass 
distributions in general for the cosmic geometry  
and take into account that the observed clustering may not
be a property of the supporting set but a property of the mass distribution on it. 
This aspect of clustering is directly related to the nature of voids: in a fractal {\em set}, the set of voids is simply the complementary set, which is totally empty, 
whereas the voids in 
a mass distribution with full support have tiny mass concentrations inside (see 
Figs.~\ref{devil} and \ref{G-homeo}). The latter option describes better our 
present knowledge of cosmic voids \cite{I4,Gott,voids}.

In conclusion, the appropriate geometrical setting for the study of the large-scale structure 
of the universe is the geometry of mass distributions, as a generalization of the 
geometry of sets.
Naturally, a generic mass distributions is more complex in three-dimensional space 
than in one-dimensional space. 
In particular, one has to consider properties such as connectivity, 
which are important for the web structure (Fig.~\ref{devil}). Since the supporting set
of a geometrical model of cosmic structure is probably trivial, namely, the full volume, 
one cannot formulate the morphological properties in terms of the topology
of that set. So it is necessary to properly {\em define} the topology of a
mass distribution. This can be done in terms of the topology of sets of particular mass 
concentrations but is beyond the scope of this work.  
The classification of mass concentrations is certainly useful, 
as a previous step and by itself. The standard way of classifying mass concentrations
constitutes the {\em multifractal analysis} of measures \cite[ch.~17]{Falcon}.

\subsection{Multifractal Analysis of mass distributions}
\label{MF}

The original idea of hierarchical clustering of galaxies 
considers galaxies as equivalent entities and constructs a sort
fractal set as the prolongation of the hierarchy towards higher scales
(up to the scale of homogeneity) and towards lower scales   
(the prolongation to lower and indeed 
infinitesimal scales is necessary to actually define a mathematical fractal set). 
The chief observable is the fractal dimension, which has been generally 
obtained through Eq.~(\ref{xi}), namely, from the value of $\g$ estimated  
through the correlation function of galaxy {\em positions} 
\cite{Mandel,Cole-Pietro,Pee,Peebles}. 
By using the galaxy positions, 
the density $\rr$ in Eq.~(\ref{xi}) is the galaxy number density 
instead of the real mass density. 
Both densities should give similar results, if the mass were distributed 
uniformly over the galaxy positions.
However, Pietronero \cite{Pietronero} noticed that galaxies are not equivalent to one another 
and that, in fact, their masses span a broad range. Therefore, he argued that 
the fractal dimension based on galaxy positions would 
not be enough and one should extend the concept of fractal to that of multifractal, 
with a spectrum of different exponents \cite{Pietronero}. 

This argument was a good motivation for considering mass distributions other 
than Mandelbrot's typical example, namely, a self-similar fractal set with mass 
uniformly distributed on it. This type of distribution, characterized by only 
one dimension, is usually called {\em unifractal} or {\em monofractal}, in contrast with the
general case of {\em multifractal} distributions, characterized by many dimensions.
However, one can have a sample of equivalent points from
an arbitrary mass distribution, as is obvious by considering this distribution as 
a probability distribution. Actually, several researchers were soon making 
multifractal analyses of the distribution of galaxy positions, 
without considering their masses, which were not available
\cite{Jones,Bal-Schaf}. This approach is not mathematically wrong, but it 
does not reveal the properties of the real mass distribution. 
We show in Sect.~\ref{MF-LSS} the result of a proper multifractal analysis 
of the real mass distribution.

Let us introduce a practical method of multifractal analysis that employs a 
coarse-grained mass distribution and is called ``coarse multifractal analysis'' 
\cite{Falcon}. 
A cube that covers the mass distribution
is divided in a lattice of cells (boxes) of edge-length $l$. Fractional
statistical moments are defined as
\begin{equation}
\M_q(l) = \sum_i \left(\frac{m_i}{M}\right)^{q}, 
\label{Mq}
\end{equation}
where the index $i$ runs over the set of {\em non-empty} cells, 
$m_i$ is the mass in the cell $i$, $M= \sum_i m_i$ is the total mass, and $q \in \mathbb{R}$.
The restriction to non-empty cells is superfluous for $q>0$ but crucial for $q\leq 0$.
The power-law behavior of $\M_q(l)$ for $l\ra 0$ is given by the exponent $\tau(q)$ such that
\begin{equation}
\M_q(l) \sim l^{\tau(q)},
\label{tau}
\end{equation}
where $l$ is the box-size. The function $\tau(q)$ is a {\em global} 
measure of scaling.

On the other hand, mass concentrates on each box with a 
different ``strength'' $\a$, such that the mass in the box is $m \sim l^{\a}$
(if $l<1$, then the smaller is $\a$, the larger is $m$ and the greater is the strength). 
In the limit of vanishing $l$, the exponent $\a$ becomes a {\em local} fractal dimension. 
Points with $\a$ larger than the ambient space dimension 
are mass depletions and not mass concentrations.
It turns out that the spectrum of local dimensions is related to the function $\tau(q)$ by
$\a(q)= \tau'(q)$ \cite{Falcon}.
Besides, every set of points with common local
dimension $\a$ forms a fractal set with Hausdorff dimension $f(\a)$ 
given by the Legendre transform $f(\a) = q\,\a - \tau(q)$.

Of course, a coarse multifractal analysis
only yields an approximated $f(\a)$, which however has the same mathematical 
properties as the exact spectrum. 
The exact spectrum can be obtained by computing $f(\a)$ for a sequence of decreasing $l$ 
and establishing its convergence to a limit function.

For the application of coarse multifractal analysis to cosmology, it is 
convenient to assume that the total cube, of length $L$, 
covers a homogeneity volume, that is to say, that $L$ is the scale of transition to homogeneity.
The integral moments $\M_n(l)$, $n \in \mathbb{N}$, are connected with the 
$n$-point correlation functions, which are generally employed in cosmology 
\cite{Pee} (we can understand the non-integral moments with $q>0$ as an interpolation). 
A straightforward calculation, starting from Eq.~(\ref{Mq}), shows that 
$\M_q$, for $q>0$, can be expressed in terms of the 
coarse-grained mass density as
\begin{equation}
\M_q(l) = \frac{\langle \rr^q\rangle_l}{\langle \rr\rangle^q}\left(\frac{l}{L}\right)^{3(q-1)},
\label{Mqr}
\end{equation}
where 
$\langle \rr^q\rangle_l$ is the average corresponding to coarse-graining length~$l$.
In particular, $\M_2$ is connected with the density two-point correlation function,  
introduced in Sect.~\ref{FG}.
In fact, 
we can use Eq.~(\ref{xi}) to integrate $\xi$ in a cell and calculate $\langle \rr^2\rangle_l\,.$
Assuming that $\xi \gg 1$, we obtain 
$$
\frac{\langle \rr^2\rangle_l}{\langle \rr\rangle^2} \sim  \left(\frac{L}{l}\right)^{\g}.$$
Therefore, $\M_2 \sim l^{3-\g}$, and, comparing with Eq.~(\ref{tau}), $\tau(2) = 3-\g$. 
In a monofractal (or unifractal), this is the only dimension needed and actually
$\tau(q) = (q-1)\tau(2)$, which implies $\a(q)= \tau'(q)=\tau(2)$ and
$f(\a) = q\,\a - \tau(q)=\tau(2)$. In this case, we can write Eq.~(\ref{tau}) as
\begin{equation}
\M_q \sim \M_2^{q-1}.
\label{hierar}
\end{equation}
This relation, restricted to integral moments, 
is known in cosmology as a hierarchical relation of moments \cite{Pee,Borga}.
Actually, if all $m_i$ are equal for some $l$ in Eq.~(\ref{Mq}), then $\M_q = \M_0^{1-q}$, 
for all $q$, and therefore $\M_q = \M_2^{q-1}$.
Let us notice that, when $\langle \rr^q\rangle_l \simeq \langle \rr\rangle^q$, Eq.~(\ref{Mqr}) 
implies that $\tau(q) = 3(q-1)$, that is to say, we have a ``monofractal'' of dimension 3.
In cosmology, this happens in the homogeneous regime, namely, for $l \ra L$, 
but not for $l \ra 0$.

\begin{figure}[H]
\centering
\includegraphics[width=9.5cm]{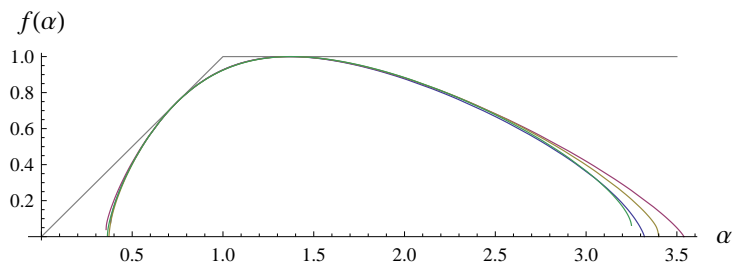}
\caption{Several coarse multifractal spectra for the 
random mass distribution of Fig.~\ref{G-homeo}, showing very good convergence. The limit 
function $f(\a)$ is a typical example of multifractal spectrum of self-similar mass distributions.}
\label{spec_G}
\end{figure}   

The multifractal spectrum $f(\a)$ provides a classification of mass concentrations,
and $f(\a)$ is easy to compute through Eq.~(\ref{Mq}) and (\ref{tau}), followed by a Legendre transform. Its computation is equally simple for mass distributions in one or 
several dimensions. Actually, we know what type of multifractal spectrum to expect for 
self-similar multifractals: 
it spans an interval $[\a_{\mathrm{min}},\a_{\mathrm{max}}]$, is concave (from
below), and fulfills $f(\a) \leq \a$ \cite{Falcon}. 
Furthermore, the equality $f(\a)=\a$ is
reached at one point, which represents  
the set of singular mass concentrations that 
contains the bulk of the mass and is called the ``mass concentrate.''
The maximum value of $f(\a)$ is the dimension of the support of the mass distribution.
We deduce that
$f(\a)$ is tangent to the diagonal segment from $(0,0)$ to $(\mathrm{max} f,\mathrm{max} f)$ 
and to the horizontal half line that starts at $\a=\mathrm{max} f$
and extends to the right. 
The values of $q=f'(\a)$ at these two points of tangency are 1 and 0, respectively. 
Beyond the maximum of $f(\a)$, we have $q<0$.
Notice that the restriction to $q>0$ and, in particular, to the calculation of only 
integral moments, misses information on the structure corresponding to 
a considerable range of mass concentrations (possibly, mass depletions). 
An example of this type of multifractal spectrum, with $\mathrm{max} f=1$,  
is shown in Fig.~\ref{spec_G} (a full explanation of this figure is given below).

The maximum of $f(\a)$, as the dimension of the support of the mass distribution, 
provides important information about voids \cite{voids}. If $\mathrm{max} f$ equals 
the dimension of the ambient space, then either voids are not totally empty or there is 
a sequence of totally empty voids with sizes that decrease so rapidly that the
sequence does not constitute
a {\em fractal hierarchy} of voids. This type of sequence is characterized by
the Zipf law or Pareto distribution \cite{Mandel,voids0,voids}. 
It seems that $\mathrm{max} f=3$ in cosmology and, furthermore, 
that there are actually no totally empty voids, as discussed below. 
In mathematical terms, one says that the mass distribution 
has full support, and we can also say that it is {\em non-lacunar}. 
We have already seen examples of non-lacunar distributions 
in Sect.~\ref{FG}, which are further explained below. Let us point out a general property
of non-lacunar distributions. 
When $\mathrm{max} f$ equals 
the dimension of the ambient space, the corresponding $\a$ is larger, because $\a \geq f(\a)$ 
(it could be equal, for a non-fractal set of points, but we discard this possibility). 
The points with such values of $\a$ are most abundant [$f(\a)$ is maximum] and also are
mass depletions that belong to non-empty voids. 
The full set of points in these voids
is highly complex, so it is not easy to define individual voids with 
simple shapes \cite{voids} and may be better to speak of 
``the structure of voids'' \cite{Gott}. 

As simple examples of multifractal spectrum calculations, 
we consider the calculations for the one-dimensional mass 
distributions studied in Sect.~\ref{FG}.
The case of inverse devil's staircases is special, because the local dimension of  
{\em zero-size} mass condensations is $\a=0$. Furthermore, 
the set of locations of mass condensations is countable and hence 
has zero Hausdorff dimension, that is to say, $f(0)=0$
(in accord with the general condition $f(\a) \leq \a$).
The complementary set, namely, the void region, has Hausdorff dimension $f(\a)=1$ and 
local dimension $\a = 1/h >1$ [deduced from the step length scaling, Eq.~(\ref{gap-law})]. 
Since the set of mass concentrations fulfills $f(\a) = \a$, for $\a=0$, it is in fact the 
``mass concentrate.'' The closure of this set is the full interval, 
which is the support of the distribution and is dominated by voids that 
are not totally empty, a characteristic property of non-lacunar distributions. 
This type of multifractal spectrum, with just one point 
on the diagonal segment from $(0,0)$ to $(1,1)$ and another on the horizontal half line
from $(1,1)$ to the right, has been called 
{\em bifractal} \cite{Aurell-Frisch-N-B}.
However, the multifractal spectrum that is calculated through the $\tau$ function
involves the Legendre transform and, therefore, is 
the convex hull of those two points, namely,
the full segment from $(0,0)$ to $(1/h,1)$.
The extra points find their meaning in a coarse-grained approach to the bifractal.

Let us comment further on the notion of bifractal distribution. We first recall that 
a bifractal 
distribution of galaxies was proposed by Balian and Schaeffer \cite{Bal-Schaf}, 
with the argument that there is one fractal dimension for clusters of galaxies and 
another for void regions. In fact, 
the bifractal proposed by Balian and Schaeffer \cite{Bal-Schaf}
is more general than the one just studied, because they assume that the clusters, 
that is to say, the mass concentrations, do not have $f(\a) = \a=0$ 
but a positive value. 
Such a bifractal can be considered as just a simplification of a standard 
multifractal, because 
any multifractal spectrum has a bifractal approximation. Indeed, any multifractal spectrum 
contains two distinguished points, as seen above; namely, the point where $f(\a)=\a$,
which corresponds to the mass concentrate, and the point where $f(\a)$ is maximum.  
The former is more important regarding the mass and the second is more important regarding 
the supporting set of the mass distribution (the set of ``positions''), 
whose dimension is $\mathrm{max} f$. 
However, this supporting set is riddled by voids in non-lacunar multifractals. 


The second example of multifractal spectrum calculation corresponds to the 
one-dimensional random mass distribution generated for Fig.~\ref{G-homeo}.
This mass distribution is just a mathematical construction that is not connected with physics
but is a typical example of continuous strictly singular mass distribution.
The multifractal spectrum is calculable analytically, but we have just employed coarse multifractal analysis of the particular realization in Fig.~\ref{G-homeo}. 
This is a good exercise to assess the convergence of the method for a mass distribution 
with arbitrary resolution. 
In Fig.~\ref{spec_G} are displayed four coarse spectra, for 
$l=2^{-13},2^{-14},2^{-15}$ and $2^{-16}$, which show perfect
convergence in the full range of $\a$, except at the ends, especially,
at the large-$\a$ end, corresponding to the emptiest voids.
The limit is a standard spectrum of a self-similar multifractal, unlike 
the bifractal spectrum of the inverse devil's staircase. 
Since the mass distribution of Fig.~\ref{G-homeo} is non-lacunar, 
$\mathrm{max} f$ reaches its highest possible value, $\mathrm{max} f=1$.
The long range of values with $\a>1$ suggests a rich structure of voids.

Finally, let us generalize to three dimensions
the multifractal spectrum of the inverse devil's staircase 
generated by the one-dimensional adhesion model, for a later use.  
In three dimensions, in addition to 
point-like masses, there are filaments and sheets. 
Therefore, the (concave envelop of the) graph of $f(\a)$ 
is the union of the diagonal segment from $(0,0)$ to $(2,2)$ and the 
segment from $(2,2)$ to $(\a_{\mathrm{max}},3)$, with some $\a_{\mathrm{max}}>3$
for voids, as mass depletions.%
\footnote{The conjecture that, along a given line, there is essentially 
a one-dimensional situation \cite{V-Frisch} suggests that $\a_{\mathrm{max}}=2+1/h$.}
There are three distinguished points of the multifractal spectrum 
in the diagonal segment, namely, $(0,0)$, $(1,1)$ to $(2,2)$, corresponding
to, respectively, nodes, filaments and sheets, where mass concentrates.
Each of these entities corresponds to a Dirac-delta singularity in the mass density:
nodes are simple three-dimensional Dirac-delta distributions, while filaments or sheets 
are Dirac-delta distributions on one or two-dimensional topological manifolds.
There is an infinite but countable number of singularities of each type. Moreover, 
the set of locations of each type of singularities is dense, which is equivalent to 
saying that every ball, however small, contains singularities of each type. 
The mathematical description is perhaps clumsy but the 
intuitive geometrical notion can be grasped just by observing Fig.~\ref{devil} 
(it actually plots a two-dimensional example, but the three-dimensional case is 
analogous).

\subsection{Multifractal Analysis of The Large-Scale Structure}
\label{MF-LSS}

Of course, the resolutions of the available data of large-scale structure are 
not as good as in the above algorithmic examples, 
but we can employ as well  
the method of coarse multifractal analysis with shrinking coarse-graining length. 
As regards quality, we have to distinguish two 
types of data: data from galaxy surveys and data from cosmological $N$-body simulations. 

\begin{figure}[H]
\centering
\includegraphics[width=7.6cm]{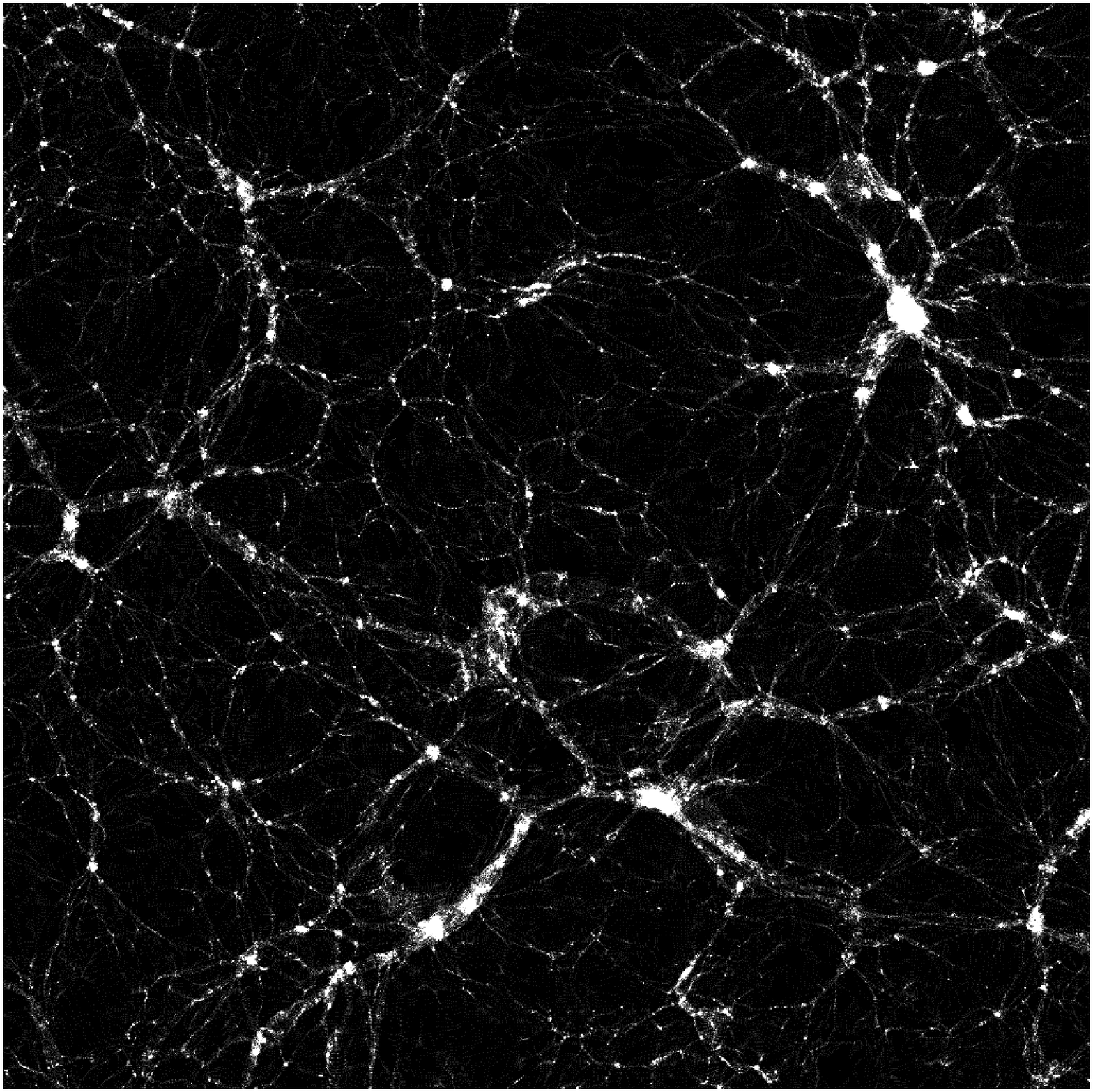}
\caption{Slice of $125\,h^{-1}\mathrm{Mpc}$ side length of the 
cosmic web structure 
formed in the {\em Bolshoi} cosmological $N$-body simulation.
}
\label{Bolshoi-cw}
\end{figure}   

The first multifractal analyses of large-scale structure employed old
galaxy catalogues. These catalogues contained galaxy positions but no galaxy masses. 
Nevertheless, Pietronero and collaborators \cite{Cole-Pietro,Sylos-PR}
managed to assign masses heuristically and carry out 
a proper multifractal analysis, with Eq.~(\ref{Mq}) and (\ref{tau}).
The results were not very reliable, because the range of scales available was insufficient. This problem also affected contemporary multifractal analyses that only employed 
galaxy positions. 
Recent catalogues contain more faint galaxies and, therefore, longer scale ranges.
Furthermore, it is now possible to make reasonable estimates of the masses of galaxies.
In this new situation, we have employed the rich Sloan Digital Sky Survey and
carried out a multifractal analysis of the galaxy distribution in the data release 7,  
taking into account the (stellar) masses of galaxies \cite{I-SDSS}.
The results are reasonable and are discussed below.
At any rate, galaxy survey data are still poor for a really thorough 
multifractal analysis.

\begin{figure}
\centering
\includegraphics[width=6.8cm]{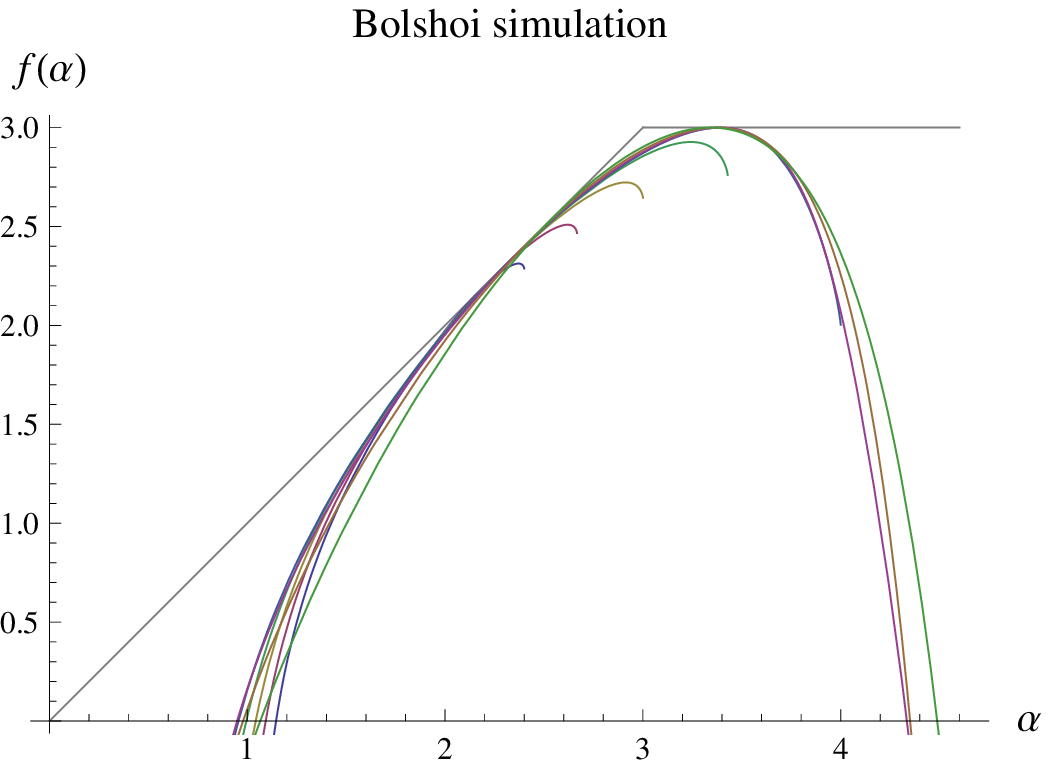}\hspace{2mm}
\includegraphics[width=7.8cm]{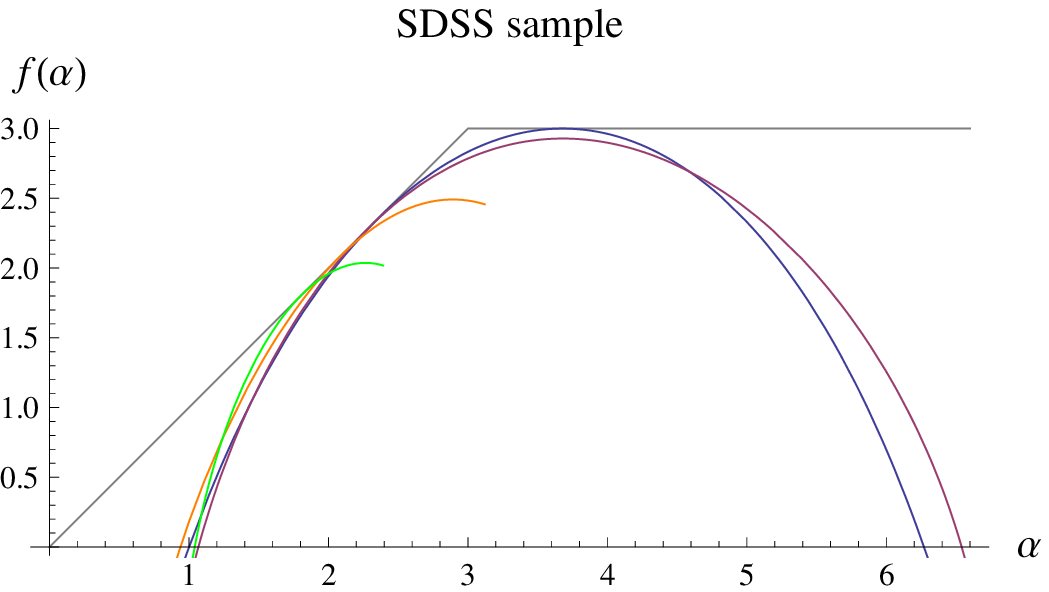}
\caption{The multifractal spectra of the dark matter distribution in the Bolshoi simulation 
(coarse-graining lengths $l=3.91, 1.95, 0.98, 0.49, 0.24, 0.122, 0.061, 0.031
\;\mathrm{Mpc}/h$)
and of the stellar mass in a volume-limited galaxy sample of the Sloan Digital Sky Survey
(coarse-graining lengths $l=
11.5, 5.76, 2.88, 1.44\;\mathrm{Mpc}/h$).}
\label{Bol+SDSS}
\end{figure}

Fortunately, the data from cosmological $N$-body simulations are much better.
On the one hand, state-of-the-art simulations handle billions of particles, so they 
afford excellent mass resolution. Several recent simulations provide a 
relatively good scaling range, namely, more than two decades (in length factor), 
whereas it is hard to get even one decade in galaxy redshift surveys.
Of course, the scale range is still small if compared to the range in our example of
random mass distribution in one dimension,
but it is sufficient, as will be shown momentarily. 
On the other hand, the $N$ bodies simulate the full
{\em dark matter} dynamics, whereas galaxy surveys are restricted to {\em stellar matter}, 
which only gives the distribution of baryonic matter (at best).
Of course, the distribution of baryonic matter can be studied 
in its own right. A comparison between multifractal spectra of 
large scale structure in dark matter $N$-body simulations and in galaxy surveys 
is made in Ref.~\cite{I-SDSS}, using recent data 
(the {\em Bolshoi} simulation, Fig.~\ref{Bolshoi-cw}, and the Sloan Digital Sky Survey).
We show in Fig.~\ref{Bol+SDSS} the shape of the respective multifractal spectra.

Both the spectra of Fig.~\ref{Bol+SDSS} are computed with coarse multifractal analysis, so 
each plot displays several coarse spectra. The plot in the left-hand side 
corresponds to the Bolshoi simulation, which 
has very good mass resolution:  
it contains $N=2048^3$ particles in a volume of $(250\,\mathrm{Mpc})^3/h^3$. 
The coarse spectra in the left-hand side of Fig.~\ref{Bol+SDSS} correspond 
to coarse-graining length $l=3.9$ Mpc/$h$ and seven subsequently halved scales.
Other cosmological $N$-body simulations yield a similar result and, in particular, one 
always finds convergence of several coarse spectra to a limit function. 
Since the coarse spectra converge 
to a common spectrum in different simulations, 
we can consider the found multifractal spectrum of dark matter reliable.
Notice however that the convergence is less compelling for larger $\a$,
corresponding to voids and, specifically, to emptier voids. 

In the right-hand side of Fig.~\ref{Bol+SDSS} is the plot of four 
coarse spectra of a volume-limited sample (VLS1) from the 7th data release of the 
Sloan Digital Sky Survey, with $1765$ galaxies in 
a volume of $3.68\cdot 10^4\;\mathrm{Mpc}^3/h^3$, for coarse-graining volumes 
$v = 1.53\cdot10^{3},\, 191.,\,23.9,\,2.99\;\mathrm{Mpc}^3/h^3$
(the coarse-graining length can be estimated as $l=v^{1/3}$).
Four coarse spectra is the maximum number available for any sample that we have analyzed
\cite{I-SDSS}. These four coarse spectra converge in the zone $\a < 2$, 
of relatively strong mass concentrations, but there are fewer converging spectra
for larger $\a$ and the shape of the spectrum of mass depletions ($\a > 3$) is questionable. 
While it seems that the multifractal spectrum reaches its 
maximum value possible, $\mathrm{max} f = 3$, 
so that the distribution can be non-lacunar, 
the values $f(\a)$ for larger $\a$ are quite uncertain \cite{I-SDSS}. 
In particular, it is unreliable that $\a_\mathrm{max} \gtrsim 6$, a quite large value, 
to be compared with 
the considerably smaller and more reliable value of $\a_\mathrm{max}$ derived from 
$N$-body simulations.

At any rate, we can assert a good concordance between 
the multifractal geometry of the cosmic structure in cosmological $N$-body simulations 
and galaxy surveys, to the extent that the available data allow us to test it, 
that is to say, 
in the important part of the multifractal spectrum up to its maximum (the part such that 
$q>0$).
The common features found in Ref.~\cite{I-SDSS} and visible in Fig.~\ref{Bol+SDSS} are:
(i) a minimum singularity strength $\a_\mathrm{min} = 1$; 
(ii) a ``supercluster set'' of dimension $\a=f(\a)\simeq 2.5$ where the mass concentrates; 
and (iii) a non-lacunar structure (without totally empty voids).
It is to be remarked that $\a_\mathrm{min} = 1$ corresponds to the edge of 
diverging gravitational potential. 

As regards the visual morphological features, 
a web structure can be observed in both the $N$-body and 
galaxy distributions. In particular, the slice of the 
{\em Bolshoi} simulation displayed in Fig.~\ref{Bolshoi-cw} definitely 
shows this structure and, furthermore, it  
looks like the two-dimensional distribution from the adhesion model displayed in 
the right-hand side of Fig.~\ref{devil}. 
However, when we compare 
the real multifractal spectrum of the large-scale structure with 
the multifractal spectrum of 
the three-dimensional mass distribution formed according to the adhesion model, 
as seen at the end of Sect.~\ref{MF}, we can appreciate considerable differences.
Therefore, in spite of the fact that the real cosmic web {\em looks like} the web 
structure generated by 
the adhesion model, the differences in their respective multifractal spectra 
reveal differences in the distributions. Indeed, 
the cosmic web multifractal spectrum, well approximated by the left-hand side graph 
in Fig.~\ref{Bol+SDSS}, contains no point-like or filamentary Dirac-delta distributions 
and it does not seem to contain sheet distributions, although the graph passes close 
to the point $(2,2)$. The differences can actually be perceived by carefully 
comparing the right-hand side of Fig.~\ref{devil} with Fig.~\ref{Bolshoi-cw}.
Notice that the features in the adhesion-model cosmic web appear to be sharper, 
because they correspond to zero-size singularities, namely, 
nodes having zero area and 
filaments zero transversal length. 
In Fig.~\ref{Bolshoi-cw}, 
point-like or filamentary mass concentrations are not so sharp, as 
corresponds to power-law concentrations.

In fact, the notion of gravitational sticking in the adhesion model is a rough 
approximation that can only be valid for sheets, because the accretion of matter to a
point ($\a=0$) or a sharp filament ($\a=1$) involves the dissipation of 
an infinite amount of energy (in Newtonian gravity). It can be proved that
any mass concentration with $\a < 1$ generates a diverging 
gravitational potential.
Therefore, it is natural that the real cosmic web multifractal spectrum
starts at $\a=1$ (Fig.~\ref{Bol+SDSS}). 
Moreover, Fig.~\ref{Bol+SDSS} shows that
$f(1)=0$, that is to say, that the set of locations of singularities with 
$\a=1$ has dimension zero, which suggests that  
the gravitational potential is finite everywhere.

The above considerations are based on the assumption of continuous matter, and even 
the adhesion model initially assumes a fluid model for the dynamics, 
even though it gives rise to point-like mass condensations (at nodes). However,
$N$-body dynamics is intrinsically discrete and this can modify the 
type of mass distribution on small scales. In fact, cosmological $N$-body simulation
have popularized {\em halo models}, in which the matter distribution on small 
scales is constructed on a different basis.

\subsection{Halo Models}
\label{halos}

The idea of galactic dark matter halos has its roots in the study of the dynamics 
in the outskirts of galaxies, which actually motivated
the introduction of the concept of dark matter. These dark halos are to be related 
to the visible shapes of galaxies but should be, in general, more spherical and extend 
to large radii, partially filling the intergalactic space and even including
several galaxies. Therefore, 
the large-scale structure of dark matter should be the combination of  
quasi-spherical distributions within halos with the distribution of halo centers.
This picture is analogous to a statistical model of the galaxy distribution 
introduced by Neyman and Scott in 1952 \cite{NeySc}, 
in which the full distribution of galaxies 
is defined in terms of 
clusters of points around separate centers 
combined with a distribution of cluster centers.
Neyman and Scott considered a ``quasi-uniform'' distribution of cluster centers 
and referred to the Poisson distribution \cite{NeySc}. Of course, 
modern halo models consider the clustering of halo centers \cite{CooSh}.

The halo model is apparently very different from the models that we have seen above, 
but it is actually connected with the basic fractal model and its cluster hierarchy. 
Indeed, we can consider Neyman and Scott's original model as consisting of
a hierarchy with just two levels, namely, point-like galaxies and their clusters, 
with little correlation between cluster centers.  
A specific formulation of such model is Fry's hierarchical Poisson model, which he employed 
to study the statistics of voids \cite{Fry}. This two-level hierarchy
can be generalized by adding more levels (as already discussed \cite{voids}). 
An infinite cluster hierarchy satisfying Mandelbrot's
conditional cosmological principle constitutes the basic fractal model.
Of course, halo models are normally defined so that they have patterns of clustering 
of halos that are independent of the distributions inside halos, departing 
from self-similarity.

In fact, the distribution of halo centers is often assumed to be 
weakly clustered and is sometimes treated perturbatively. 
However, it tends to be recognized that halos must preferentially lie in sheets 
or filaments and actually be ellipsoidal \cite{Bol,Mona}.
Neyman and Scott required the distribution of points around the center  
to be spherical, without further specification. 
Modern halo models study in detail the halo {\em density profiles}, in addition to 
deviations from spherical symmetry. 
For theoretical reasons and also relying on 
the results of cosmological $N$-body simulation, density profiles are assumed 
to have power-law singularities at their centers. There is a number 
of analytic forms of singular profiles, 
with parameters that can be fixed by theory or fits \cite{CooSh}.
A detailed account of this subject is beyond the scope of this work. 
It is especially important, in our context, the 
power-law singularity and its exponent, which normally 
has values in the range $[-1.5,-1]$. 

We can compare the density singularities in halos  
with the singular mass concentrations already studied in multifractal distributions.
A coarse-grained multifractal distribution indeed
consists of a collection of discrete mass concentrations and can be formulated as
a fractal distribution of halos \cite{fhalos,I4}. Models based on this idea constitute 
a synthesis of fractal and halo models. 
Furthermore, the idea naturally agrees with the current tendency to 
place halos along filaments or sheets and hence assume 
strong correlations between halo centers.
The range $[-1.5,-1]$ of the density power-law exponent is equivalent to the 
range $\a \in [1.5,2]$ of the mass concentration dimension, defined by $m \sim r^\a$.
In view of Fig.~\ref{Bol+SDSS}, this range is below the value 
for the mass concentration set [fulfilling $f(\a_\mathrm{con}) = \a_\mathrm{con}$], which is
$\a_\mathrm{con} \simeq 2.5$.
This difference can be attributed to a bias towards 
quasi-spherical rather than ellipsoidal halos, because the former are more concentrated.
In the web structure generated by the adhesion model,  
nodes are more concentrated than filaments or sheets.

It is to be remarked that the theories of halo density profiles are consistent with 
a broad range of power-law exponents and even with no singularity at all 
(Gurevich-Zybin's theory \cite{GZ} is an exception). In fact, the quoted density-singularity 
range is mainly obtained from the analysis of $N$-body simulations. For this and other
reasons, it is necessary to judge the reliability of $N$-body simulation in the range of scales 
where halos appear \cite{I-Galax}. Notice that the reliability of 
$N$-body simulations on small scales 
has been questioned for a long time \cite{Melott,Splinter,JMB}.
The problem is that two-body scattering spoils the collisionless dynamics, 
altering the clustering properties.
Although the mass distribution inside an $N$-body halo is smooth (except at the center), 
this can be a consequence of discreteness effects: 
an $N$-body halo experiences a transition 
from a smooth distribution to a very anisotropic and non-smooth 
web structure over a scale that should
vanish in the continuum limit $N\ra\infty$ \cite{Bol}. 

To conclude this section, let us try to present an overview of
the prospects of the halo model. The original idea of the halo model, namely, 
the description of the cosmic structure in terms of smooth halos with centers 
distributed in a simple manner and, preferably, almost uniformly, 
seems somewhat naive, inasmuch as it is an {\em ad-hoc} combination of 
two simple types of distributions, which can only be valid on either 
very small or very large scales.
Moreover, {\em smooth} halos seem to be the result of $N$-body simulations 
in an unreliable range of scales. 
In spite of these problems, the notion of galactic dark matter halos, that is to say, 
of small-scale dark matter distributions that control the dynamics of 
baryonic matter and the formation of galaxies, is surely productive.
However, the cold-dark-matter model has problems of its own, because the collisionless
dynamics tends, anyway, to generate too much structure on small scales, and 
it is expected that the baryonic physics will change the dynamics on those scales 
\cite{CDM_PNAS}.
The halo model provides a useful framework to study this question \cite{Popo}.
The ongoing efforts to unify the halo and multifractal models 
\cite{fhalos,I4,MN,ChaCa,Bol}
will hopefully help to give rise to a better model of 
the large-scale structure of the universe.

\section{Formation of the Cosmic Web}
\label{form}

The halo or fractal models are of descriptive nature, although they can be employed 
as frameworks for theories of structure formation. In contrast, the Zeldovich approximation
and the adhesion model {\em constitute} a theory of structure formation. Unfortunately, this 
theory is quite simplistic, insofar as the nonlinear gravitational dynamics is 
very simplified, to the extent of being apparently absent. 
At any rate, in the correct equations, the equation for the gravitational field is linear 
and the nonlinearity is of hydrodynamic type (see the equations in Sect.~\ref{cosmo-dyn} below).
This nonlinearity gives rise to turbulence and, in the adhesion model in particular, 
to Burgers turbulence, which is highly nontrivial. 

So we place our focus on the properties of Burgers turbulence.  
However, we begin by surveying other models of the formation of 
the cosmic web that are 
not directly based on approximations of the cosmological equations of motion 
(Sect.~\ref{voids}). Next, we formulate these equations in Sect.~\ref{cosmo-dyn} and
proceed to the study of turbulence
in Burgers dynamics, first by itself (Sect.~\ref{B-turb}) and
second in the setting of stochastic dynamics (Sect.~\ref{KPZ}). Finally, we consider 
the full nonlinear gravitational dynamics (Sect.~\ref{chaos}).

\subsection{Models of formation of cosmic voids}
\label{voids}

Let us recall two models of the formation of 
the cosmic web, namely, 
the {\em Voronoi foam} model \cite{Sahni-C,Rien}
and Sheth and van de Weygaert's model of the hierarchical evolution of voids
\cite{Sheth-vdW}. Both these models focus on the formation and evolution of cosmic voids, 
but the Voronoi foam model is of geometric nature whereas the hierarchical void model is 
of statistical nature. 
They are both heuristic and 
are not derived form the dynamical equations 
of structure formation, but the Voronoi foam model can be 
naturally connected with the adhesion model.

The Voronoi foam model focuses on the underdense zones 
of the primordial 
distribution of small density perturbations. The minima of the density field 
are the peaks of the primordial gravitational potential and, therefore, the seeds of voids. Indeed, voids form as matter flows away from those points. 
The basic model assumes that the centers of void expansion are random, namely, that 
they form a Poisson point field, and assumes that every center generates a Voronoi cell. 
The formation of these cells can be 
explained by a simple {\em kinematic} model that 
prescribes that matter expands at the same velocity from adjacent centers and, therefore, a 
``pancake'' forms at the middle, that is to say, a condensation forms in a region of 
the plane perpendicular to the joining segment through its midpoint \cite[Fig.~40]{Rien}. 
The Voronoi foam results from the simultaneous condensation on all these regions. 

Let us describe in more detail the formation of the cosmic web according to this 
kinematic model \cite{Sahni-C,Rien}. After the matter condenses on a wall between 
two Voronoi cells, it continues to move within it and away from the two nuclei, until it 
encounters, in a given direction, 
the matter moving along two other walls that belong to the cell of a third nuclei. 
The intersection of the three walls is a Voronoi edge, on which the matter from 
the three walls condense and continues its motion away from the three nuclei. Of course, 
the motion along the edge eventually leads to an encounter with 
the matter moving along three other edges that belong to the cell of a fourth nuclei, and 
it condenses on the corresponding node. 
All this process indeed follows the rules of the adhesion model, but the 
Zeldovich approximation of the dynamics is replaced by a simpler kinematic model, 
which produces a particular cosmic web, namely, a Voronoi foam.
It is easy to simulate matter distributions that form in this way, and those  
distributions are suitable for a correlation analysis \cite{Rien}. 
This analysis reveals that the two-point correlation function is a power law, with 
an exponent $\g$ that grows with time, as matter condenses on lower dimensional elements 
of the Voronoi foam. Naturally, $\g$ is close to two in some epoch (or close 
to the standard value $\g=1.8$).

The only parameter of the Voronoi foam model with random Voronoi nuclei 
is the number density of these nuclei. 
In fact, this construction is well known in stochastic geometry and is known as 
the Poisson-Voronoi tessellation. 
The number density of nuclei determines the average size of Voronoi cells, 
that is to say, the average size of cosmic voids. The exact form of the 
distribution function of Voronoi cell sizes is not known, but it is well approximated 
by a gamma distribution that is peaked about its average \cite{Kiang}. 
Therefore, the Voronoi foam model reproduces early ideas about cosmic voids, 
namely, the existence of a ``cellular structure'' with a limited range of cell sizes 
\cite{cellular}. 
This breaks the self-similarity of the cosmic web found with the adhesion model 
by Vergassola {\em el al} \cite{V-Frisch} (see Sect.~\ref{ss_web}).   
The self-similarity of cosmic voids, in particular, is demonstrated by 
the analysis of galaxy samples or cosmological $N$-body simulations \cite{voids}. 
Notice that 
gamma distributions of sizes 
are typical of voids of various shapes in a Poisson point field  
\cite{voids}. In a Voronoi foam 
with random nuclei, the matter points would be the nodes of condensation, namely, 
the Voronoi vertices, and these do not constitute a Poisson point field but are 
somehow clustered. At any rate, a distribution of sizes of cosmic voids 
that is peaked about its average is not realistic.

Moreover, a realistic foam model
must account for the clustering of minima of the primordial density field 
and for the different rates of void expansion, giving rise to other types 
of tessellations. In fact, such models seem 
to converge towards the adhesion model \cite{Sahni-C}. 
The adhesion model actually produces a Voronoi-like tessellation, characterized by 
the properties of the initial velocity field (or the initial density field)
\cite{Bernardeau-Vala}. In cosmology, 
the primordial density field is not smooth and, hence, has no isolated 
minima. Although a smoothened or coarse-grained field can be suitable for studying the 
expansion of cosmic voids, the distribution of voids sizes is very different 
from the one predicted by the basic Voronoi foam model \cite{voids}.

Let us now turn to Sheth and van de Weygaert's hierarchical model of evolution of voids 
\cite{Sheth-vdW}. It is based on 
an ingenious reversal of the Press-Schechter and {\em excursion set} approaches to 
hierarchical gravitational clustering. 
In fact, the idea of the Press-Schechter approach is better suited for the evolution 
of voids than for the evolution of mass concentrations (a comment on the original approach 
can be found at the end of Sect.~\ref{chaos}). The reason is the ``bubble theorem'', 
which shows that aspherical underdense regions tend to become spherical in time, unlike 
overdense regions \cite{Sahni-C,Rien}.

Sheth and van de Weygaert's hierarchical model of voids is more elaborate than 
the Voronoi foam model but it also predicts a characteristic void size. 
To be precise, it predicts that the
size distribution of voids, at any given time, is peaked about a characteristic void size,  
although the evolution of the distribution is self similar. 
Actually, the lower limit to the range of void sizes is due to a small-scale cut-off,  
which is put to prevent the presence of voids inside overdense regions that are supposed to collapse (the ``void-in-cloud process'') \cite{Sheth-vdW}. 
The resulting distribution of void sizes looks like the gamma distribution of 
the Voronoi foam model, and Sheth and van de Weygaert \cite{Sheth-vdW} indeed argue that 
the hierarchical model, which is more elaborate, somehow justifies the 
Voronoi foam model, which is more heuristic. 

If the small-scale cut-off is ignored, by admitting that voids can form inside overdense regions, then the distribution of void sizes in Sheth and van de Weygaert's model has 
no characteristic void size and, for small sizes, it is 
proportional to the power $-1/2$ of the void volume \cite{Sheth-vdW}. 
Power-law distributions of void sizes 
characterize {\em lacunar} fractal distributions, which follow the Zipf law for voids, 
but the exponent is not universal and depends on the fractal dimension 
\cite{voids}. At any rate, the results of analyses 
of cosmic voids and the cosmic web itself,  
presented in Ref.~\cite{voids} and Sect.~\ref{MF-LSS},  
show that the matter distribution is surely multifractal 
and {\em non-lacunar}. This implies that cosmic voids are not totally empty but have 
structure inside, so that the self-similar structure does not manifest itself 
in the distributions of void sizes but in the combination of matter concentrations and 
voids (as discussed in \cite{voids}). In fact, the cosmic web generated by the 
adhesion model with scale-invariant initial conditions is a good example of this type 
of self similarity (Sect.~\ref{ss_web}).

Therefore, the conclusion of our brief study of models of formation of cosmic voids is
that just the adhesion model is more adequate than these other models for describing 
the formation of the real 
cosmic web structure. The adhesion model derives from the Zeldovich approximation to 
the cosmological dynamics, which we now review.

\subsection{Cosmological dynamics}
\label{cosmo-dyn}

We now recall the cosmological equations of motion in the Newtonian limit, 
applicable on scales small compared to the Hubble length and away
from strong gravitational fields \cite{Pee,Peebles,Sahni-C}.
The Newtonian equation of motion of a test particle in an expanding background
is best expressed in terms of the comoving coordinate ${\bm x} = {\bm r}/a(t)$
and  the {\em peculiar} velocity ${\bm u} = \dot{\bm x} = {\bm v} - H {\bm r}$,
where $H= \dot{a}/a$ is the Hubble constant (for simplicity, we can take $a=1$ 
in the present time). 
So the Newton equation of motion, $d{\bm v}/{d t} = {\bm g}_\mathrm{T}$,
with total gravitational field ${\bm g}_\mathrm{T}$, can be written as  
\begin{equation}
\frac{d{\bm u}}{d t} + H {\bm u} = {\bm g},
\label{Newton-eq}
\end{equation}
where the net gravitational field,
${\bm g} = {\bm g}_\mathrm{T} - {\bm g}_\mathrm{b}$, is 
defined by subtracting the background
field ${\bm g}_\mathrm{b}=(\dot{H} + H^2){\bm r}$. The equation for this background
field is 
$$3(\dot{H} + H^2) = \nabla\cdot{\bm g}_\mathrm{b} = -4\pi G\rr_\mathrm{b} + \L c^2,$$
that is to  say, 
the dynamical FLRW equation for pressureless matter (with a cosmological constant). 
This equation gives $H(t)$ and hence $a(t)$. 
To have a closed system of equations, we must add to Eq.~(\ref{Newton-eq}) the equation
that gives ${\bm g}$ in terms of the density fluctuations, 
\begin{equation}
\nabla\cdot{\bm g} = -4\pi G(\rr-\rr_\mathrm{b}),
\label{Newton-eq1}
\end{equation}
and the continuity equation,  
\begin{equation}
\frac{\partial \rr}{\partial t} + 3H\rr + \nabla\cdot(\rr {\bm u}) = 0.
\label{cont-eq}
\end{equation}
These three equations form a closed nonlinear system of equations.
 
The equations can be linearized when and where the 
density fluctuations are small. Within this approximation, 
one actually needs only Eq.~(\ref{Newton-eq}), 
which is nonlinear because of the convective term 
${\bm u}\cdot \nabla{\bm u}$ in $d{\bm u}/{d t}$ but becomes 
linear when the density fluctuations and hence ${\bm g}$ and ${\bm u}$ 
are small. 
The linearized motion is 
\begin{equation}
{\bm x} = {\bm x}_0 + b(t)\,{\bm g}({\bm x}_0),
\label{Lag-map}
\end{equation}
where $b(t)$ is the growth rate of linear density fluctuations. 
Redefining time as
$\tilde{t}=b(t)$, we have a simple motion along straight lines, with constant velocities 
given by the initial peculiar gravitational field.

\subsection{Burgers Turbulence}
\label{B-turb}

The Zeldovich approximation prolongs the linear motion, Eq.~(\ref{Lag-map}),
into the nonlinear regime, where ${\bm u}$ is not small. 
To prevent multi-streaming, the adhesion model
supplements the linear motion with a {\em viscosity} term, 
as explained in Sect.~\ref{ss_web}, resulting in the equation
\begin{equation}
\frac{d \widetilde{\bm u}}{d \tilde{t}} \equiv
\frac{\partial \widetilde{\bm u}}{\partial \tilde{t}} + 
\widetilde{\bm u}\cdot \nabla\widetilde{\bm u} =
\nu \nabla^2\widetilde{\bm u}, 
\label{Burg}
\end{equation}
where $\widetilde{\bm u}$ is the peculiar velocity in $\tilde{t}$-time.
To this equation, it must be added the condition of no vorticity (potential flow), 
$\nabla \times \widetilde{\bm u} = 0$, implied by $\nabla \times {\bm g}({\bm
  x}_0) = 0$. 
It is to be remarked that the viscosity term is in no way fundamental and the same 
job is done by any dissipation term, in particular, by a higher-order 
{\em hyperviscosity} term \cite{Boyd}.
Eq.~(\ref{Burg}) is the (three-dimensional) Burgers equation for 
pressureless fluids (tildes are suppressed henceforth). 
Any $\nu > 0$ prevents multi-streaming and
the limit ${\nu \ra 0}$ 
is the high Reynolds-number limit, which 
gives rise to Burgers turbulence. Whereas incompressible turbulence is
associated with the development of vorticity, Burgers turbulence 
is associated with the development of {\em shock fronts}, namely, 
discontinuities of the velocity. 
These discontinuities arise at caustics and give rise to matter accumulation 
by inelastic collision of fluid elements.
The viscosity $\nu$ measures the thickness of shock fronts,
which become true singularities in the limit ${\nu \ra 0}$. 

A fundamental property of the Burgers equation is that it is integrable; namely, 
it becomes a linear equation by applying to it the Hopf-Cole transformation 
\cite{Frisch-Bec,Bec-K,Shan-Zel}. 
This property shows that the Burgers equation is a very special case of the Navier-Stokes
equation. 
In spite of it, Burgers turbulence is a useful model of turbulence in the 
irrotational motion of very compressible fluids.
At any rate, the existence of an explicit solution of the Burgers equation 
makes the development of Burgers turbulence totally dependent on the properties of
the initial velocity distribution, in an explicit form. 
The simplest solutions to study describe the formation of isolated 
shocks 
in an initially smooth velocity field. 
Fully turbulent solutions are provided by a
self-similar initial velocity field, such that 
$\bm{u}(\l \bm{x}) = \l^h \bm{u}(\bm{x}),\; h\in (0,1)$,
as advanced in Sect.~\ref{ss_web}.
Indeed, Eq.~(\ref{Burg}) is scale invariant in the limit $\nu\ra 0$, namely, it
is invariant under simultaneous space and time scalings, $\l \bm{x}$, $\l^{1-h} {t}$, 
for arbitrary $h$, if the velocity is scaled as $\l^{h}
\bm{u}$, and therefore has self-similar solutions, such that
\begin{equation}
\bm{u}(\bm{x}, {t} ) =  {t}^{h/(1-h)} \bm{u}\!\left(\bm{x}/ {t}^{1/(1-h)},1\right).
\label{dyn-sim}
\end{equation}
A self-similar velocity field that is fractional Brownian
on scales larger than the ``coalescence length'' $L \propto {t}^{1/(1-h)}$
contains a distribution of shocks that is self-similar on scales below $L$. 
Such velocity field constitutes a state 
of {\em decaying turbulence}, since kinetic energy is 
continuously dissipated in shocks \cite{Frisch-Bec,Bec-K,Shan-Zel}.
In the full space $\mathbb{R}^3$, as is adequate for the cosmological setting, 
the transfer of energy from large scales, where the initial conditions hold, to 
the nonlinear small scales, proceeds indefinitely.

The simplicity of this type of self-similar turbulence allows one to analytically prove 
some properties and make reasonable conjectures about others \cite{Frisch-Bec,Bec-K,V-Frisch}.
A case that is especially suitable for a full analytical treatment is 
the one-dimensional dynamics with $h=1/2$, namely, with Brownian initial velocity. 
As regards structure formation, the dynamics 
consists in the {\em merging} of smaller structures to form larger ones, 
in such a way that the size of the largest structures is 
determined by the ``coalescence length'' and, therefore, they have masses that grow 
as a power of ${t}$ \cite{Shan-Zel,V-Frisch}. The merging process agrees with 
the ``bottom-up'' picture for the growth of cosmological structure \cite{Peebles}.
This picture is appropriate for the standard cold dark matter cosmology.

From the point of view of the theory of turbulence, centered on the 
properties of the velocity field, the formed structure consists of 
a dense distribution of shock fronts with very variable magnitude. 
Therefore, the kinetic energy dissipation takes place in an extremely non-uniform manner, 
giving rise to {\em intermittency} \cite{Shan-Zel}.
A nice introduction to the phenomenon of intermittency 
as a further development of Kolmogorov's theory of turbulence can be found in 
Frisch's book \cite[ch.~8]{Frisch}. 
Kolmogorov's theory assumes that the rate of specific dissipation of energy 
is constant.
Intermittency is analyzed by modeling how kinetic energy is distributed among
structures, either vortices in incompressible turbulence
or shocks in irrotational compressible turbulence. 
Mandelbrot proposed that intermittency is the consequence 
of dissipation in a fractal set rather than uniformly \cite[\S 10]{Mandel}. 
Mandelbrot's idea is embodied in the $\b$-model of the energy cascade \cite{Frisch}. 
However, this model is monofractal and a bifractal model is more adequate, 
especially, for Burgers turbulence. 
Although this bifractality refers to 
the velocity field \cite{Frisch}, it corresponds, in fact, to the bifractality of the 
mass distribution in the one-dimensional adhesion model already seen in Sect.~\ref{MF}. 
Let us briefly study the geometry of intermittency in Burgers turbulence.

Kolmogorov's theory of turbulence builds on 
the Richardson cascade model and is based on {\em universality}
assumptions, valid in the limit of infinite Reynolds number (${\nu \ra 0}$)
and away from flow boundaries.
As explained by Frisch \cite{Frisch}, it is possible to deduce,
by assuming homogeneity, isotropy, and finiteness of energy dissipation in the 
limit of infinite Reynolds number, 
the following scaling law for the moments of longitudinal velocity increments 
\begin{equation}
\langle \left(\d \bm{u} \cdot \bm{r}/r \right)^k \rangle \propto (\e r)^{k/3},
\label{K}
\end{equation}
where $$\d \bm{u}(\bm{x}) = \bm{u}(\bm{x}+\bm{r}/2) -\bm{u}(\bm{x}-\bm{r}/2),$$
$\e$ is the specific dissipation rate, 
and $k \in \mathbb{N}$. 
To introduce the effect of intermittency, the Kolmogorov scaling law can be generalized to
\begin{equation}
\langle \left|\d \bm{u}\right|^q \rangle = A_q\, r^{\z(q)}, 
\label{gK}
\end{equation}
where $q \in \mathbb{R}$, and $A_q$ does not depend on $r$ (and does not have to 
be related to $\e$). Eq.~(\ref{gK}) expresses that the fractional statistical 
moments of $\left|\d \bm{u}\right|$ have a power-law behavior given by the 
exponent $\z(q)$. Therefore, the equation is analogous to the combination of 
Eqs.~(\ref{Mq}) and (\ref{tau}) for multifractal behavior. 
The functions $\z(q)$ and $\tau(q)$ 
are indeed connected in Burgers turbulence (see below).

If $\z(q) \propto q$ in Eq.~(\ref{gK}), as occurs in Eq.~(\ref{K}), then 
the probability $P(\d \bm{u})$ is Gaussian.
The initial velocity field that we have chosen for Burgers turbulence 
is Gaussian, namely, fractional Brownian, with 
$\langle \left|\d \bm{u}\right|^2 \rangle \propto r^{2h}$ (Sect.~\ref{ss_web}).
Therefore, 
$\z(q) = hq$ if $r \gg L({t})$ (which would fit Eq.~(\ref{K}) for $h=1/3$).
In general, intermittency 
manifests itself as concavity of $\z(q)$ \cite{Frisch}. 
In the Burgers turbulence generated by the given initial conditions,
a type of especially strong intermittency develops for any ${t} > 0$ and $r \ll L({t})$.
This  intermittency is such that 
$\z(q)$,  in addition to being concave, has a maximum value equal to one. 
This can be proved by generalizing
the known argument for isolated shocks \cite{Frisch-Bec}
to a dense distribution of shocks, as is now explained.

Let us replace in Eq.~(\ref{gK}) 
the ensemble average with a spatial average (notice that 
fractional statistical moments of $\d m/M$ in Eq.~(\ref{Mq}) are 
defined in terms of a spatial average).
The average in Eq.~(\ref{gK}), understood as a spatial average, can be 
split into regular and singular points (with shocks). 
At regular points, the inverse Lagrangian map $\bm{x}_0(\bm{x},{t})$ is well defined and, 
from Eq.~(\ref{Lag-map}),
$$\bm{u}(\bm{x},{t}) = \bm{u}_0(\bm{x}_0) = \bm{g}_0(\bm{x}_0) = \frac{\bm{x}-\bm{x}_0(\bm{x},{t})}{{t}}\,.$$
Self-similarity in time allows one to set ${t}=1$ without loss of generality.
Hence, we deduce the form of the spatial velocity increment over $\bm{r}$ at a regular point
$\bm{x}$, 
\begin{equation}
\d \bm{u}(\bm{x}) = \bm{r} - \d\bm{x}_0(\bm{x}).
\label{du}
\end{equation}
Since the properties of the Lagrangian map are known, one can deduce properties of the
velocity increment. For example, at a regular point $\bm{x}$,
the term $\d\bm{x}_0(\bm{x})$ is subleading for $r \ll L$, 
because the derivatives of the inverse Lagrangian map vanish. 

To proceed, we restrict ourselves to one dimension, 
in which the analysis is simpler and we have, for $r \ll L$,
$$\langle \left|\d {u}\right|^q \rangle \sim 
C_q \,r^{q} + \sum_n  \left|\d {u}_n\right|^q r,$$ 
where the first term is due to the regular points
and the second term is due to
the set of ${u}$-discontinuities (shocks) such that $\left|\d {u}_n\right| \gg r$. 
Clearly, putting ${t}$ back, $\d {u} \approx r/{t} > 0$ corresponds to the
self-similar expansion of voids, whereas $\d {u} < 0$ and
$\left|\d {u}\right| \gg r/{t}$ corresponds to shocks. 
If $q<1$, the first term dominates as $r\ra 0$, and vice versa.
Therefore, $\z(q) = q$, for $q\leq 1$, and $\z(q) = 1$, for $q>1$.
Notice that the number of terms in the sum increases as $r\ra 0$, but the series 
is convergent for $q\geq 1$, due to elementary properties of devil's staircases: 
if the spatial average is calculated over the interval 
$\D x$ that corresponds to an initial interval $\D x_0 \gg L$, 
then $\sum_{n=1}^\infty  \left|\d {u}_n\right| = \D x_0$ and 
$\sum_{n=1}^\infty  (\left|\d {u}_n\right|/\D x_0)^q < 1$ when $q> 1$.%
\footnote{Actually, the series converges for $q>D_\mathrm{BT}$, 
where $D_\mathrm{BT}$ is the Besicovitch-Taylor
exponent of the sequence of gaps, which is in this case equal to $h$, because
of the gap law $N(M>m) \propto m^{-h}$ 
\cite[p.~359]{Mandel}.
}

The form of $\z(q)$, with its maximum value of one,  
implies that the probability $P(\d \bm{u})$ is very non-Gaussian for $r \ll L$, 
as is well studied \cite{Frisch-Bec,Bec-K} and we now show. 
In one dimension, the ratio $\d {u}/r$ is a sort of ``coarse-grained derivative'', 
related by Eq.~(\ref{du}) 
to the coarse-grained density $\d x_0/r$ (the length of the 
initial interval $\d x_0$ that is mapped to the length $r$ is the coarse-grained density 
normalized to the initial uniform density).
Hence, we derive a simple relation between the density 
and the derivative of the velocity, namely, $\rr = 1 - \p_x u$. 
We can also 
obtain the probability of velocity 
gradients from $P(\d {u})$, in the limit $r\ra 0$.  However, this limit is singular and, 
indeed, the distribution of velocity gradients can be called strictly singular,
like the distribution of densities, because,  
at regular points, where $\rr=0$, then $\p_x u=1$, and at shock positions, 
where $\rr=\infty$, then $\p_x u=-\infty$. 
Naturally, it is the shocks that make $P(\d {u})$ very non-Gaussian and 
forbid that $\z(q)$ grows beyond $\z=1$. 
%
The intermittency is so strong that the form of $\z(q)$ is very different from Kolmogorov's
Eq.~(\ref{K}) and the {\em average} dissipation rate $\e$ seems to play no role. 

As the mass distribution can be expressed in terms of the velocity field, 
the intermittency of the latter is equivalent to the multifractality of the former. 
Indeed, the bifractal nature of $\tau(q)$, in one dimension, is equivalent to 
the dual form of $\z(q)$ for $q>1$ or $q \leq 1$. Let us notice, 
however, that $\z(q)$ is universal, whereas the bifractal  
$f(\a)$ studied in Sect.~\ref{MF} depends on the value of $h$ (at the point for voids), 
and so does $\tau(q)$. 
Actually, the full form of $P(\d {u})$ also depends on $h$.  

For the application to cosmology,
we must generalize these results to three dimensions. 
Although this generalization is conceptually straightforward, it brings in 
some complications: in addition to point-like mass condensations, there appear 
other two types of mass condensations, namely, filaments and sheets, the latter 
actually corresponding to primary shocks.%
\footnote{Furthermore, there is mass flow {\em inside} the higher-dimensional 
singularities, and this flow must be determined \cite{Bernardeau-Vala}.} 
The form of $\z(q)$ is still 
determined by the primary shocks and is, therefore, unaltered \cite{Frisch-Bec,Bec-K}. 
Nevertheless, the mass distribution itself is no longer bifractal, as is obvious.
The form of $f(\a)$ has already been seen in Sect.~\ref{MF}; namely, it is 
the union of the diagonal segment from $(0,0)$ to $(2,2)$ and the 
segment from $(2,2)$ to $(\a_{\mathrm{max}},3)$, with $\a_{\mathrm{max}}>3$.

In summary, the adhesion model, with initial velocity and density fields having
Gaussian power-law fluctuations, 
namely, being fractional Brownian fields, 
leads to a self-similar and strongly intermittent Burgers turbulence,  
such that various quantities of interest can be calculated analytically. 
This dynamics gives rise to a self-similar cosmic web that is quite realistic. 
Therefore, it constitutes
an appealing model of structure formation in cosmology \cite{Shan-Zel,V-Frisch}. 
However, it has obvious shortcomings: (i) in cosmology, the initial but ``processed'' 
power spectrum is not a power law \cite{Peebles,Sahni-C}; 
(ii) the Burgers equation is integrable
and therefore the adhesion model may be too simplified a
model of the gravitational dynamics, which actually is {\em chaotic}; and
(iii) the condensation of matter in shock fronts is considered as an inelastic
collision but the dissipated kinetic energy is assumed to disappear without further effect.  
In general, chaos and dissipation are connected: chaotic dynamics erases memory of initial
conditions, giving rise to an irreversible process in which entropy grows. 
To deal with these problems, we need a more general framework. A first step is taken 
by connecting the Burgers equation with a well studied stochastic equation, namely, 
the Kardar–Parisi–Zhang (KPZ) equation, as we study next.

\subsection{The Stochastic Burgers Equation}
\label{KPZ}

After studying 
freely decaying Burgers turbulence in the adhesion model,
we proceed to the study of 
{\em forced} Burgers turbulence 
\cite{Frisch-Bec,Bec-K}. It is understood that the forcing in cosmology is not external 
but is just the feedback of dissipated energy. Indeed, 
we intend to achieve a more complete understanding of the dynamics 
that is based on a general relation between dissipation and fluctuating forces, 
such as the relation embodied in the 
classical fluctuation-dissipation theorem \cite{Reif}. 
So we rely on the theory of stochastic dynamics, 
with origin in the Langevin equation for Brownian motion. The Langevin equation 
is derived by simply adding
a fluctuating force $F(t)$ to Newton's second law. This force is due to the 
interaction with the environment and must contain a 
slowly varying dissipative part and a rapidly varying part with null average (a ``noise'').

The Langevin equation has a limited scope. 
A qualitative development in stochastic dynamics consisted in considering 
partial differential equations, with fluctuating forces that depend on space and time. 
A simple example is the Edwards-Wilkinson equation, which results from adding noise
to the diffusion equation \cite{H-H--Z}. Assuming that the scalar field of 
the Edwards-Wilkinson equation is the ``height'' of some substrate, 
the equation can be employed in the study of stochastic growth. 
The Edwards-Wilkinson equation is linear 
but a more suitable equation for stochastic growth,
namely, the Kardar–Parisi–Zhang (KPZ) equation,  
was obtained by adding to it a nonlinear term.
For several reasons, this equation has become a paradigm in the theory of 
stochastic nonlinear partial differential equations. In fact, by taking the 
scalar field in the KPZ equation as a velocity potential, the equation becomes equivalent 
to the stochastic Burgers equation, in which a noise term, that is to say, a random force, 
is added to the
right-hand side of Eq.~(\ref{Burg}), so that it contains dissipation plus noise.

The presence of noise in the stochastic Burgers equation
alters the evolution of the velocity and density fields. For example, 
mass condensations can gain energy and hence fragment, so 
adhesion is no longer irreversible. It may even not happen at all.
The different kind of evolution induced by noise is best appreciated if 
the spatial region is bounded, with size $L$ (typically, the use of 
periodic boundary conditions makes space toroidal). Then, 
the total mass and kinetic energy are bounded and the free decay of
Burgers turbulence eventually leads to 
a simple state. For example, in one dimension, all the kinetic energy 
is dissipated and all the mass condensates on a single point (for generic 
initial conditions) \cite{Frisch-Bec,Bec-K}.
As noise injects energy, the dissipative dynamics cannot relax to the 
state of minimum energy but to a fluctuating state, which may or may not have 
mass condensations, according to the characteristics of the noise. 

In the Langevin dynamics, the particle motion relaxes until it is in thermal equilibrium 
with the medium and has the Maxwell velocity distribution \cite{Reif}.
The strength of the noise is given by the temperature $T$, and the velocity 
fluctuations fulfill
\begin{equation}
\langle \left[\bm{v}(t+\D t)-\bm{v}(t)\right]^2 \rangle = 
\frac{2T}{m}\,\frac{\D t}{t_\mathrm{rel}}\,,
\label{brown}
\end{equation}
where $m$ is the mass of the particle and $t_\mathrm{rel}$ is the relaxation time. 
This equation says that the variance of $\D v$ is independent of the absolute 
time $t$, and this
holds for all $t$, 
provided that $\D t \ll t_\mathrm{rel}$. 
In contrast, the value of $\langle \bm{v}(t) \rangle$ explicitly depends on $t$ 
and on the initial condition $\bm{v}(0)$.
The proportionality of the variance of $\D v$ to $\D t$
is characteristic of Brownian motion. 

The thermal noise in the Langevin equation is Gaussian and macroscopically uncorrelated 
\cite{Reif}; 
that is to say, it has {\em white} frequency spectrum. 
However, there is no reason to assume thermal equilibrium, in general. 
Indeed, various ``colors'' are considered in stochastic dynamics. 
In the theory of stochastic partial differential equations 
the noise is Gaussian and has a general power spectrum $D(k,\o)$. 
In particular, it is favored a noise with power-law spectrum of spatial correlations 
but still uncorrelated in time, namely, with $D(k) \propto k^{-2\rho}$ (see below).

The Edwards-Wilkinson equation for a scalar 
field $\phi(\bm{x},t)$ is equivalent to the stochastic Burgers equation 
for $\bm{u}=-\nabla\phi$ with the convective term suppressed.
The solution of the Edwards-Wilkinson equation 
is more difficult than the solution of the Langevin equation, 
but it is a linear equation and hence soluble, by decoupling of
the Fourier spatial modes. 
The dynamics also consists in a relaxation to a stationary state 
\cite{H-H--Z}. 
However, the relaxation takes place at a rate that depends on the 
``roughness'' of $\phi(\bm{x},t)$ at $t=0$. Indeed, Fourier spatial modes have relaxation 
times inversely proportional to $k^2$, so relaxation is faster on the 
smaller scales.

The KPZ dynamics 
also begins to relax on the small scales. On these scales, 
the nonlinear term, which corresponds to the convective term 
in the Burgers equation, becomes important.
As said above, the combination of nonlinearity and dissipation tends to build structure 
whereas the noise tends to destroy it.
In fact, the KPZ dynamics is ruled by the competition between nonlinearity and noise. 
To proceed, let us introduce the {\em dynamic scaling hypothesis} \cite{H-H--Z},
as a generalization of Eq.~(\ref{brown}). 

Given a general stochastic partial differential equation for a scalar 
field $\phi(\bm{x},t)$, it is reasonable to assume that
\begin{equation}
\langle \left[\phi(\bm{x} + \D \bm{x},t + \D t)-\phi(\bm{x},t)\right]^2 \rangle = 
\left|\D\bm{x}\right|^{2\chi} 
f\left(\D t/\left|\D\bm{x}\right|^z\right),
\label{dyn-scal}
\end{equation}
where $\chi$ and $z$ are critical exponents and $f$ is a scaling function. This function 
expresses 
the crossover between the power laws of purely spatial and purely temporal correlations:
$f(s)$ approaches a constant for $s \ra 0$
and behaves as $s^{2\chi/z}$ for $s \ra \infty$.
Eq.~(\ref{dyn-scal}) generalizes Eq.~(\ref{brown}) to both space and time and, hence, is 
analogous to the space-time similarity already considered in Eq.~(\ref{dyn-sim})
[the redefinition $f(s) = s^{2\chi/z} g(s)$ makes the connection
with Eq.~(\ref{dyn-sim}) or Eq.~(\ref{brown}) more apparent].%
\footnote{Let us notice that the averages considered in this section are taken with 
respect to realizations of the random noise, 
unlike in the preceding section, in which they are taken with respect to realizations of the  random initial conditions.} 
Eq.~(\ref{dyn-sim}) implies that the initial conditions are modified  
below a scale that grows as a power of time [with exponent $1/(1-h)$] and, 
likewise, Eq.~(\ref{dyn-scal}) implies that relaxation to the noise-induced asymptotic
distribution takes place below a scale that grows as a power of $\D t$, 
with exponent $1/z$. 


In the case of the Edwards-Wilkinson equation in $d$ dimensions for a field $\phi(\bm{x},t)$,
dynamic scaling is borne out by the explicit solution. In the solution,  
the Fourier mode $\phi_{{\bf k}}(t)$ is a linear combination of the noise Fourier modes, 
so a Gaussian noise produces a Gaussian $\phi$ \cite{H-H--Z}. This field fulfills 
Eq.~(\ref{dyn-scal}), so $\phi$ and $\bm{u}=-\nabla\phi$, for fixed $t$, are actually 
{\em fractional Brownian} fields with Hurst exponents $\chi$ and $\chi-1$, respectively. 
%
With noise spectrum $D(k) \propto k^{-2\rho}$, 
the explicit solution yields $\chi=1-d/2+\rho$ and $z=2$,  
as can also be deduced from the scaling properties of the equation. 
Therefore, $\phi$ and $\bm{u}$, at fixed $t$, become less rough as $\rho$ grows. 
For example, in one dimension and with white noise, $\chi=1/2$, so $\phi(x,t)$ describes, 
for fixed $t$, a Brownian curve, whereas $u = -\p_x\phi$ is uncorrelated (white noise like).
Moreover, the dynamics consists in a relaxation to thermal equilibrium, in which 
fluid elements acquire the Maxwell velocity distribution, in accord with
the fluctuation-dissipation theorem. 

The KPZ equation is nonlinear and its solution is a non-Gaussian field.  
Therefore, Eq.~(\ref{dyn-scal}) is just an assumption, 
supported by theoretical arguments and computer simulations.
The one-dimensional case, with white noise, is special, 
insofar as the effect of the nonlinear term is limited,
the fluctuation-dissipation theorem holds,
and $\chi=1/2$, like in the Edwards-Wilkinson equation. For general dimension and 
noise uncorrelated in time, the Burgers equation has Galilean 
invariance  (invariance under addition of a constant vector to 
$\bm{v}$ plus the consequent change of $\bm{x}$), which 
enforces the relation $\chi + z = 2$ \cite{H-H--Z}. 
In one dimension and with white noise, this relation implies that $z=3/2$, which 
shows that the KPZ and Edwards-Wilkinson equations differ in temporal correlations, 
that is to say, that the convective term of the stochastic Burgers equation does have 
an effect. Nevertheless, the dynamics also consists in a relaxation to 
the Maxwell velocity distribution. In the state of thermal equilibrium, there is no structure.

At any rate, for cosmology, we must consider the three-dimensional KPZ equation.
The study of the KPZ equation with ``colored'' noise 
has been carried out by Medina et al \cite{Medina}, 
employing {\em renormalized perturbation theory}, namely, 
perturbation of the linear equation by the nonlinear term and subsequent renormalization.  
Within this approach, the types of noise that give rise to dynamical scaling are found as 
fixed points of the {\em dynamical renormalization group}.
%
%
%
For small $k$, 
the noise power spectrum assumes the universal form 
$$D(k) \approx D_0 + D \,k^{-2\rho},\; \rho >0,$$ 
that is to say, the noise consists of a white noise plus a
power-law long-range noise \cite{Medina}.
Unfortunately, in three dimensions, renormalized perturbation theory is of little consequence: 
the only stable renormalization group fixed point
is the trivial one (vanishing nonlinear convective term) and only if $\rho<1/2$
\cite{Medina}.
This means that a noise with sufficient power on large scales inevitably leads
to a strong-coupling stationary state and 
questions the application of perturbation theory to 
the stochastic Burgers equation with spatially correlated noise in three dimensions.  
In spite of it, the {\em stochastic adhesion model} 
has been long studied in cosmology and mostly with 
perturbative treatments, which are still being applied \cite{Rigo}.
Certainly, non-perturbative methods seem more sensible \cite{I-EPL}. 

In fact, renormalized perturbation theory, in any dimension $d$, is valid only 
for $\rho<(d+1)/2$ \cite{Medina}. 
For larger $\rho$, the long-range noise correlations induce analogous 
long-range correlations in $\phi$ and $\bm{u}$ and give rise to intermittency of $\bm{u}$ 
(as studied for $d=1$ in Ref.~\cite{Verma}).
%
%
In particular, 
the field $\bm{u}$ 
becomes strongly non-Gaussian. 
This is a purely nonlinear effect and may seem counterintuitive, because, 
in the Edwards-Wilkinson equation, $\phi$ and $\bm{u}$, at fixed $t$, 
become less rough as $\rho$ grows. 
The progressive smoothness of $\bm{u}$ turns it into a real hydrodynamical field 
(in contrast with the white noise like velocity in $d=1$ with $\rho=0$, for example).   
Although the nonlinearity, namely, the convective term of 
the Burgers equation, does not spoil this smoothing, generally speaking, it
gives rise to shocks and, hence, intermittency. 
The combination of smoothing and formation of shocks can be seen, for example, 
in simulations in $d=1$ \cite[Fig.~4]{Verma}. 
Therefore, for $\rho>(d+1)/2$, we have the regime in which structures can form in spite of 
the noise, that is to say, the properly turbulent regime. 
As we focus on the field $\bm{u}$ in this regime, 
we replace Eq.~(\ref{dyn-scal}) with Eq.~(\ref{gK}) [we have $\z(2) = 2(\chi-1)$]. 

The turbulent regime of forced Burgers turbulence has been well studied, 
with a combination of computer simulations and 
theoretical arguments related to the Kolmogorov theory.
A simple argument of Chekhlov and Yakhot \cite{Che-Ya}, for $d=1$, shows that
$\rho = 3/2$ corresponds to an ``almost constant'' (logarithmic) energy flux
in Fourier space (a balanced Richardson energy cascade). 
Furthermore, their numerical simulations show that $u$ develops shocks (with the 
corresponding mass condensations) which give rise to power-law correlations with 
exponents $z=2/3$ and $\z(2)/2=\chi-1=1/3$. The latter corresponds to 
the Kolmogorov scaling, in accord with Eq.~(\ref{K}).
Let us also notice that the noise exponent $\rho = 3/2$ is such that 
the noise strength $D$ has the dimensions of the 
dissipation rate $\e$. But Chekhlov and Yakhot's argument is 
incomplete, so that larger values of $\rho$ are suitable \cite{Verma}.
A breakthrough in the study of intermittency and Kolmogorov scaling
in forced Burgers turbulence (in one dimension) 
has been the use by Polyakov \cite{Polyakov}
of non-perturbative methods borrowed from quantum field theory. 
Polyakov \cite{Polyakov} assumes that the Burgers noise correlation function 
(in ordinary space) is twice differentiable, that is to say, that $\rho \geq 5/2$.
Boldyrev \cite{Boldyrev} has extended Polyakov's methods to the range $\rho \in [3/2,5/2]$.  

For cosmology, we must consider $d=3$.
Kolmogorov's universality and its consequences for cosmology are considered as 
a guiding principle in Ref.~\cite{I-EPL}. The conclusion is that the
interesting range of the exponent $\rho$ for the cosmic structure, in $d=3$, 
is $\rho \in(5/2,7/2)$.
The lower limit $\rho=5/2$ is
such that the noise strength $D$ has the dimensions of $\e$ and the Burgers
noise correlation function is proportional to $\log r$. Moreover, 
$\e$ diverges for $\nu\ra 0$ if $\rho \leq 5/2$.
The reason for the upper limit $\rho = 7/2$ is that 
the $r$-dependent part of the noise correlation function 
depends explicitly on the system size $L$ for $\rho > 7/2$. 
The range $\rho \in(5/2,7/2)$ corresponds to large-scale forcing that is such that 
the dissipation rate $\e$ only depends on what happens on large scales and such that
the $r$-dependent part of the noise
correlation function, proportional to $r^{2\rho-5}$, is {\em universal}.
%

Furthermore, there is intermittency for $\rho >(d+1)/2 =2$, in particular, 
in the range $\rho \in(5/2,7/2)$.
To measure it, we can use Eq.~(\ref{gK}), where now 
the exponent $\z(q)$ depends on the noise exponent $\rho$
(and the average is with respect to the random noise). 
The intermittency increases with increasing $\rho$, and progressively affects 
exponents $\z(q)$ with lower values of $q$ \cite{HJ}.
The analysis of the range $\rho \in(5/2,7/2)$ obtains strong intermittency, namely,
$\z(q)=1$ for $q \geq 3$, but $2/3<\z(2)\leq 1$. 
In particular, $\z(2) = 4(\rho-2)/3$ while in the range $\rho \in(5/2,11/4)$ \cite{HJ}. 
At $\rho = 11/4$, the exponent $\z(2)$ reaches its maximum $\z(2)=1$, and it remains fixed 
for $\rho > 11/4$.
The minimum value $\z(2)=2/3$, for $\rho=5/2$, corresponds to the Kolmogorov scaling, 
Eq.~(\ref{K}). 

Especially interesting for cosmology is that the density two-point correlation function 
can be obtained in terms of correlation functions of the velocity field; in fact, 
it can be obtained in terms of just the two-point correlation function \cite{I-EPL}. 
Furthermore, the density two-point correlation function
is a power-law, taking the form in Eq.~(\ref{xi}), with an exponent $\gamma$ determined 
by $\z(2)$, namely, $\gamma = 2-\z(2)$.
Given the range of $\z(2)$, we deduce that 
$$1<\gamma<4/3,$$ 
which is a reasonable range (although the preferred value of $\gamma$ is somewhat larger).

To summarize, 
the stochastic Burgers equation with spatially correlated noise constitutes an 
interesting model of structure formation, in which the cosmic web is not determined 
by ad-hoc initial conditions but is the result of the interplay of 
the inelastic gravitational condensation of matter with the consequent dynamical 
fluctuations, so that energy can be conserved on the average.
The relaxation to the asymptotically stable state takes place over a scale that 
grows with time, namely, $L \propto {t}^{1/z}$, where $z$ is determined by 
the dynamics instead of the initial conditions. 
The asymptotically stable state has simple scaling properties and,
furthermore, it achieves Kolmogorov's universality. However, this universality does not lead 
to uniqueness, in the sense of total independence of the large scales: indeed, there is 
an allowed range for the noise exponent $\rho$. At any rate, the model predicts a 
{\em stable} fractal cosmic web, such that the exponent of the 
density two-point correlation function is within a reasonable range.
Furthermore, the formation of structures that are stable in the highly nonlinear regime 
connects, in spirit, with 
famous Peebles' {\em stable clustering} hypothesis \cite{Pee}, to be studied in 
the next section. 


The rationale for 
a fluctuating force is not restricted to the cosmological 
equations in the Zeldovich approximation. One can employ the original equations,
namely, Eqs.~(\ref{Newton-eq}), (\ref{Newton-eq1}) and (\ref{cont-eq}).
This system of equations can be supplemented with viscous and 
random forcing terms \cite{Antonov}. At any rate, 
it behooves us to take a more general standpoint and study the full 
dynamics of gravitational clustering.

\subsection{Nonlinearity, Chaos and Turbulence in Gravitational Clustering}
\label{chaos}

The first comprehensive attempts to treat analytically the 
theory of large scale structure formation are due to Peebles and collaborators \cite{Pee}.
They employed general principles of statistical mechanics to formulate the problem and
added a scaling hypothesis to obtain definite solutions.
This interesting work did not lead to a geometrical picture of structure formation 
until the connection with Mandelbrot's ideas \cite{Mandel} and the
development of the Zeldovich approximation and the adhesion model 
\cite{Shan-Zel}. Let us recall the basics of Peebles' approach 
(a complete introduction to it, with some novelties, 
is provided by Shani and Coles \cite{Sahni-C}).

Surely, the most general approach to the non-equilibrium statistical mechanics of 
a system of particles is based 
on the Liouville equation and the derived hierarchy of equations for phase-space probability
functions involving increasingly more particles, known as the BBGKY hierarchy 
(BBGKY stands for Bogoliubov-Born-Green-Kirkwood-Yvon). 
Every successive equation of this hierarchy 
gives the evolution of an $N$-particle correlation function but involves the 
$(N+1)$-particle correlation function. Therefore, to have a manageable set of equations, 
it is necessary to close the hierarchy at some level, by assuming some relation between 
the $N$-particle correlation function and lower order correlation functions.
Closure approaches are also much employed in the statistical theory of turbulence, and 
indeed the Kolmogorov theory can be understood as a closure approach with 
an additional scaling hypothesis \cite{Frisch}. 
Peebles and collaborators \cite{Pee} follow a similar path, considering 
density correlation functions in addition to velocity correlation functions, which 
are the only ones present in incompressible turbulence. 

In fact, the scaling hypothesis of Peebles {\em et al} focuses on the reduced two-point  
density correlation function $\xi$, 
which fulfills a simple equation that just expresses the conservation of particle pairs:
\begin{equation}
\frac{\partial \xi}{\partial t} + \frac{1}{x^2 a}
\frac{\partial}{\partial x} [x^2(1+\xi)v] = 0,
\label{pairs}
\end{equation}
where $v$ is the mean relative peculiar velocity of particle pairs.
Of course, the presence of $v$ makes this equation not closed. 
To proceed, it is necessary to assume closure relations not only 
for Eq.~(\ref{pairs}) but for the full set of equations. 
Nevertheless, it is possible to obtain information 
about $\xi$ only from the conservation of particle pairs, using 
scaling arguments \cite{Pee}.
However, these arguments are restricted to the Einstein-de Sitter cosmological model, 
with null curvature and therefore without characteristic scale, so that $\rr_\mathrm{b}(t)$ 
and $a(t)$ are power laws, namely, $\rr_\mathrm{b}(t) \propto t^{-2}$ and 
$a(t) \propto t^{2/3}$. Furthermore, the initial spectrum of density fluctuations 
must be a power law.

Given the absence of characteristic scales, Peebles and collaborators propose similarity
solutions of the equations, namely, solutions that are functions of the sole 
variable $s=x/t^\a$, with an exponent $\a$ that is to be determined. 
In particular, they seek a similarity
solution $\xi(s)$ of Eq.~(\ref{pairs}).
As this equation is not closed, one has to assume a relation between the 
expectation value of the relative velocity and $\xi$, and, furthermore, that this 
relation adopts a simple scaling form in the linear and very nonlinear regime, namely,
in the cases $\xi \ll 1$ and $\xi \gg 1$. In the linear case, one knows that
$\xi \propto t^{4/3}x^{-3-n}$, 
in accord with the Zeldovich approximation, 
$\xi \propto \tilde{t}^2 x^{-3-n}$, with $\tilde{t} \propto t^{2/3}$ 
for the Einstein-de Sitter cosmology ($n$ is the Fourier power spectrum exponent).
Therefore, one deduces that $\a=4/(3n+9)$.

Of course, the problematic case is the very nonlinear regime. To deal with it, 
Peebles and collaborators proposed the {\em stable clustering} hypothesis,
namely, that the average relative velocity of particle pairs 
vanishes in physical (not comoving)
coordinates \cite{Pee}. This hypothesis may seem natural for tightly bound
gravitational systems. It plays a role that is somewhat analogous to 
Kolmogorov's hypothesis of constant dissipation rate $\e$ in turbulence.
Indeed, as the constancy of $\e$ provides a link between large and small scales, 
so does the stable clustering hypothesis. In particular, substituting $v=-\dot{a}x$ 
in Eq.~(\ref{pairs}), and, furthermore,
taking into account that $\xi \gg 1$, it is easy to solve the equation and 
obtain $\xi(s) \propto s^{-\g}$, where $\g = 2/(\a+2/3)$. In terms of variables $(t,x)$
or $(a,r)$, where $r=ax$ is the physical distance, 
\begin{equation}
\xi \propto t^{2\a/(\a+2/3)}x^{-2/(\a+2/3)} \propto a^{3}\,r^{-2/(\a+2/3)},
\label{xi_EdS}
\end{equation}
Naturally, 
as regards the $r$-dependence, Eq.~(\ref{xi_EdS})
is just a particular case of
Eq.~(\ref{xi}), in which $\g$ is expressed in terms of $\a$, which  
can be derived from the initial power-spectrum exponent $n$ (in 
the Einstein-de Sitter cosmology), giving $\g = 2/(\a+2/3)=3(3+n)/(5+n)$.
As regards the dependence on $a$, that is to say, on time, Eq.~(\ref{xi_EdS}) 
has new content, which can be generalized.

First, let remark that Eq.~(\ref{xi_EdS}) 
can be deduced without invoking the equation for conservation of particle pairs or
closure relations, by means of a heuristic argument
also due to Peebles 
\cite[\S 22]{Peebles}. This argument is based on the estimation of the density 
of a bound system, which can be assumed to give the {\em average conditional density}, 
namely, the average density at distance $r$ from an occupied point. This 
statistic is expressed as
$$
\frac{\langle \rr(\bm{0})\,\rr(\bm{r})\rangle}{\langle \rr\rangle} =
\langle \rr\rangle \,[1+\xi(r)]\,,$$
where $\langle \rr\rangle=\rr_\mathrm{b}$ is the background cosmic density. 
For a bound system, $\xi \gg 1$, and the density of a system of size $r$ is the conditional 
density $\rr_\mathrm{b}\,\xi(r)$.
The average conditional density is especially useful to define 
the extremely nonlinear limit, in which $\xi \ra \infty$ and 
$\rr_\mathrm{b}\ra 0$ while its product stays finite. This limit actually allows one 
to dispense with the transition to homogeneity and study directly a 
fractal mass distribution \cite{Mandel,Pietronero}.

To estimate the density of a bound system, one can follow the evolution of 
an overdensity $\d \rr = \rr - \rr_\mathrm{b}$ of comoving size $x$, 
which initially grows 
as $\d \rr \propto \rr_\mathrm{b}\,t^{2/3}x^{-(3+n)/2}$. Over a time 
$t(x) \propto x^{3(3+n)/4}$, the overdensity is of the order of magnitude 
of the background density and the linear theory is no longer valid [notice that 
this time is the ``coalescence time'' that can be deduced in the adhesion model 
by inverting the ``coalescence length'' expression $L \propto \tilde{t}^{1/(1-h)}$].
In terms of the physical size $r$, the time for nonlinearity fulfills the relation 
$t \propto [r/a(t)]^{3(3+n)/4}$, which, 
solving for $t$, gives $t(r) \propto r^{3(3+n)/(10+2n)}$.
At this time, the background density is $\rr_\mathrm{b} \propto t^{-2} 
\propto r^{-3(3+n)/(5+n)}$, and our overdensity, as approximately twice the 
background density, follows the same rule. In fact, a factor of twice the 
background density seems hardly sufficient for the formation of a bound system; 
for example, the spherical collapse model favors a factor that is about 180 \cite{Sahni-C}.
The exact value of this factor is irrelevant and the important point is that 
the self-similar evolution of an overdense blob of size $r$ leads to a bound system 
with density $\propto r^{-3(3+n)/(5+n)}$. Therefore, interpreting this density as a 
conditional density, one obtains 
$$\xi \propto \rr_\mathrm{b}^{-1}\,r^{-3(3+n)/(5+n)}\propto a^{3}\,r^{-3(3+n)/(5+n)},$$
in accord with Eq.~(\ref{xi_EdS}).

We can draw more general conclusions
by only appealing to the stable clustering hypothesis and without assuming any 
scaling in the cosmological evolution. 
Indeed, if particle pairs keep, on the average, their relative physical distances, then
the average conditional density is a function $f(r)$ that does not depend on time. 
Therefore, $\xi = a^{3}f(r)$, where $f$ is an arbitrary function. 
This form of $\xi$ is also obtained as the general solution of Eq.~(\ref{pairs}) under 
stable clustering conditions, namely, $v=-\dot{a}x$ and $\xi \gg 1$, which 
allow us to write Eq.~(\ref{pairs}) as
$$
a\frac{\partial \xi}{\partial a} - \frac{1}{x^2}
\frac{\partial}{\partial x} (x^3 \xi) = 0.
$$
One can check that $\xi = a^{3}f(ax)$ solves the equation by direct substitution.

Of course, the stable clustering hypothesis can be questioned: 
Peebles says that ``merging might be expected to dissipate the clustering hierarchy 
on small scales'' \cite[\S 22]{Peebles}, and indeed the effect of merging is often
presented as an argument against the hypothesis.
However, this argument disregards that stable clustering occurs only {\em on the average},
that is to say, that merging can be accompanied by 
splitting, keeping the average physical distance of particle pairs constant. 
Actually, the conservation of energy in the local inertial frame demands 
it so. We have used this argument for introducing
the stochastic adhesion model in Sect.~\ref{KPZ}, a model in which
the merging and subsequent fragmentation of mass condensations takes place 
in a {\em stable} manner.
The stable clustering hypothesis in cosmology is actually related to the 
possibility of studying, in the local inertial frame and
without a cosmological model, the gravity laws 
and, specifically, the laws of bound systems;  
a possibility of which Newton certainly took advantage.%
\footnote{The problem of the influence of the global cosmological expansion on local dynamics 
is still being discussed \cite{Car-Giu} and is connected with another polemic issue, namely, 
back-reaction \cite{Buchert-br}.}

Accepting the stable clustering hypothesis, we can further argue that $f(r)$ is singular
as $r\ra 0$, and that it must have, in particular, a power law form $f(r)\propto r^{-\g}$.
But this form of $f$ is {\em independent} of the hypothesis. 
Notice that the average conditional density has dimensions of mass density, 
so that a finite limit of $f(r)$ in the limit $r\ra 0$ would define a universal mass density. 
However, there is no universal mass density in the theory of general relativity.%
\footnote{Adding quantum mechanics, we have the universal mass density 
$c^5/(h G^2) \simeq 10^{97}$ kg/m$^3$ ($h$ is the Planck constant). 
In this regard, a large but reasonable universal mass density would constitute 
a naturalness problem analogous to but less serious than the problem of the 
cosmological constant (or vacuum energy density) 
\cite{Weinberg}. 
Of course, reasonable universal mass densities can be obtained by adding 
other universal constants, e.g., the nucleon mass, the electron charge, etc.}
On the other hand, a universal mass density could also be given 
by the initial conditions, but the only density in them is $\rr_\mathrm{b}$, which
is time dependent and too small: it is actually the limit of 
the average conditional density 
for $r\ra \infty$. Therefore, it is natural to assume that $f(r)$ is singular
as $r\ra 0$.

A power-law singularity is not the only possibility but is the most natural 
possibility. In fact, the power-law form of $\xi$
is required by the notion of hierarchical clustering (Sect.~\ref{FG}),
which is related to the full form of Peebles' closure approach.
If we dispense with the full form and just keep the stable clustering with a
power-law form of $\xi$,
there is no way to obtain the value of the exponent $\g$, because there 
is no reason to relate it to the initial conditions, which do not need to be self-similar.
The evolution of the FLRW cosmological model does not need to be self-similar
either. Therefore, 
if $\g$ could be derived from the initial conditions, this derivation would 
present an awfully difficult problem.
A possible option that is conceptually different is that $\g$ be 
determined by a {\em nonlinear eigenvalue problem} \cite{Bar}. One example 
of the solution of a problem of this type is provided by 
Polyakov's non-perturbative theory of forced Burgers turbulence \cite{Polyakov}. Another
example, which is even more relevant, is provided by 
Gurevich and Zybin's theory \cite{GZ}, described below.

Let us remark that the presence of length scales, in the FLRW cosmological 
model or in the initial spectrum of density fluctuations, does not rule out 
a sort of space-time similarity in the strongly nonlinear regime and let us remark that 
such a situation has already been shown to arise in 
the stochastic adhesion model, in Sect.~\ref{KPZ}. 
Naturally, the zero-size mass condensations of the adhesion model are 
not realistic, 
so it is necessary to analyze what type of 
mass concentrations can form under the effect of Newton's gravity.


The formation of power-law density singularities in collisionless gravitational dynamics
has been proved by Gurevich and Zybin's analytic theory \cite{GZ}. 
This study is especially interesting, since 
it brings to the fore the role of nonlinear dynamics and chaos in the formation 
of gravitational structure.
They take an {\em isolated} overdensity of generic type, namely, 
a generic local maximum of the density field (assumed to be differentiable), and 
study its evolution. This evolution leads to 
a density singularity in a dynamical time 
and, after that, it consists in the subsequent development 
of a multistreaming flow with an increasing
number of streams that oscillate about the origin, within a region
defined by the first caustic. In this region, the multiplicity of streams in 
opposite directions gives rise to {\em strong mixing}, which leads to a steady state
distribution with an average density. This average density has a power-law
singularity at the origin, with an exponent that can be calculated (under reasonable 
assumptions). This exponent is {\em independent} of the particular form of 
the initial overdensity. 

In the case of spherical symmetry, Gurevich and Zybin find 
$$\rr(r) \propto r^{-2}\,[\ln(1/r)]^{-1/3}.$$ 
It corresponds to a diverging gravitational potential,  
but the divergence is weak and does not induce large velocities.
Let us recall from Sect.~\ref{MF-LSS} that the local 
mass concentration exponent $\a_\mathrm{min} = 1$ is found in cosmological 
$N$-body simulations and the SDSS stellar mass distribution, and notice that
it corresponds to $\rr(r) \propto r^{-2}$ in the case of spherical symmetry.
Arguably, the strongest mass concentrations have spherical symmetry.
The density $\rr(r) \propto r^{-2}$ is also the singular isothermal density profile, 
where thermal equilibrium is represented by an $r$-independent average velocity and 
the Maxwell distribution of velocities.
An $r$-independent circular velocity of stars is observed in the outskirts of 
spiral galaxies (this observation is, of course, one of the motivations for the 
existence of dark matter); so the isothermal density profile may have a role 
on these small scales \cite[\S 3]{Peebles}.
However, the collisionless gravitational dynamics does not lead to thermal equilibrium. 
In particular, the distributions of velocities in the steady states found 
by Gurevich and Zybin are very anisotropic \cite{GZ}.

For the sake of generality, Gurevich and Zybin treat the case without 
full spherical symmetry, 
that is to say, with low ellipticity, so that transverse velocities are small.
Remarkably, this case leads to a singularity that is not exactly a power law but
can be well approximated by one, 
namely, $\rr(r) \propto r^{-\b}$ with $\b \simeq 1.7$--1.9. 
Furthermore, this form is not very sensitive to the degree of ellipticity.
The opposite case, of very large ellipticity, can be approximated by one-dimensional 
or two-dimensional solutions. Gurevich and Zybin only consider one-dimensional flows
and derive $\rr(x) \propto x^{-4/7}$. In this case, it is possible to study 
in some detail the nonlinear caustic oscillations and the role of transfer 
of energy successively to higher harmonics \cite{GZ}.
The picture is analogous to the Richardson energy cascade in the theory 
of incompressible turbulence \cite{Frisch}.
As regards the one-dimensional power-law exponent 4/7, it has to be noticed that 
Aurell et al \cite{Aurell_2} analyze a soluble example of one-dimensional collapse, with 
different initial conditions, and obtain a less singular value.

In general, that is to say, in the collapse in dimensions higher than one 
and without symmetry,
it is commonly accepted that power-law singularities must appear, but  
there is no agreement about the exponent. This question is related, of course,
to the problem of halo density profiles in the halo model, commented in Sect.~\ref{halos}.
We have seen that a range of exponents has been considered, relying on various arguments and
on the results of simulations. 
Let us notice that halo models usually assume a bias 
towards quasi-spherical mass concentrations and, therefore, towards more singular values of 
the mass concentration exponent $\a$. 
Here we have to consider all singularities on the same footing.

If Gurevich and Zybin's theory were to be completed 
by adding the case of two-dimensional collapse, then one could think of 
obtaining a universal exponent
by averaging the exponents over suitable positions of singularities at nodes, filaments
and sheets of the cosmic web. 
Gurevich and Zybin indeed consider the problem of how to integrate their results 
for isolated singularities in a hierarchical structure. They divide this 
structure into an intermediate nonlinear range, set to 50--100 Mpc, where the 
cosmic web (``cellular'') structure appears and a lower-scale range, where steady 
nondissipative gravitational singularities appear \cite{GZ}.
However, this is an artificial 
restriction of the cosmic web structure, amounting to a separation into 
a regular ``cellular'' structure and a sort of halos, which are smooth, 
except for their central singularities.
In fact, a distribution of {\em isolated} and {\em smooth} halos 
is only an approximation of the real cosmic web \cite{fhalos}. The halos are part 
of the cosmic web structure rather than separate entities, and they may not be 
smooth \cite{Bol}. 

In particular,
Gurevich and Zybin's treatment of the collapse of {\em isolated} overdensities ignores 
that maxima of the initial density field are not isolated: in the
``bottom-up'' model of growth of cosmological structure, isolated 
density maxima of the initial density field only exist provided that this field
is coarse grained, which introduces an {\em artificial} scale \cite{BBKS}.
The resulting halos, with central singularities but smooth profiles, belong in
the model of fractal distribution of halos, which is just a coarse-grained
approximation of a multifractal model \cite{fhalos,Bol}.

In fractal geometry, the density is not a suitable variable, because it only takes 
the values zero or infinity, as explained in Sect.~\ref{FG}.
Besides, in a multifractal model, with a range 
of local exponents $\a$, the average of local exponents over spatial positions 
takes a special form: 
for global magnitudes such as the 
moments $\M_q$, each is dominated by a specific set of singularities, with local dimension 
$\a(q)=\tau'(q)$. 
The bulk of the mass belongs to the mass concentrate, with exponent $\a_\mathrm{con}=\a(1)$, 
but the singularities of this set are not particularly strong in the real cosmic web
($\a_\mathrm{con} \simeq 2.5$, see Fig.~\ref{Bol+SDSS}). According to the
argument, given above, 
that the density of bound systems is to be identified with the average conditional 
density and, hence, is connected with $\M_2$, we may think that the important exponent
is $\a(2)<\a(1)$. Certainly, the focus is usually placed on the exponent $\g$ in 
Eq.~(\ref{xi}), which is the one most accessible to measures and the one
that Peebles' closure approach tries to determine. However, 
the correlation dimension 
$\tau(2) = 3 - \g$ does not directly determine the corresponding local dimension $\a(2)$,
unless the multifractal spectrum is known [in general, $\a(2)<\tau(2)$].

These difficulties do not arise if we have only one fractal dimension. In fact, 
Peebles' closure approach 
only employs the three-point correlation function, in addition to the 
two-point correlation function, and assumes the hierarchical form for it \cite{Pee,Sahni-C},
which is valid for a monofractal but not for a general multifractal.
However, the present data support full multifractality 
(Fig.~\ref{Bol+SDSS}) and
so does Gurevich and Zybin's analytic theory, unless it is restricted to 
quasi-spheric mass concentrations. 
Moreover, even the intuitive image of the cosmic web, consisting of sheets, filaments 
and knots, shows that it could hardly involve only one dimension.
But the three-dimensional morphology of the cosmic web is by no means the cause of 
multifractality: one-dimensional cosmological $N$-body simulations already show 
multifractality \cite{Miller}.
It seems that we are obliged to accept that we have to deal with a 
range of fractal dimensions, namely, with a full multifractal spectrum that we cannot 
quite model yet. 

Multifractality of the mass distribution formed by gravitational 
clustering is surely a universal phenomenon, linked to the intermittency of 
highly nonlinear processes. Nonlinear gravitational clustering is indeed a 
phenomenon very related to turbulence, which began to be studied somewhat earlier 
but is in a similar state of development. 
The specific nature of gravity leads to some helpful constraints, chiefly,
the stable clustering hypothesis, but this is not sufficient for a solution and 
specific nonlinear methods are needed \cite{Sahni-C}. The guiding idea for studying
nonlinear phenomena like turbulence or gravitational clustering is surely 
the scaling symmetry, which is naturally 
motivated by the observed multifractality. The renormalization group prominently 
features among the analytic methods based on scaling, 
as has been advocated before \cite{I_RG}.
The efficacy of this method is attested by its results for the stochastic adhesion model, 
specifically, the results obtained 
employing nonperturbative formulations of the renormalization group \cite{Kloss-Canet}.
Stochastic gravitational clustering demands a random noise with an arbitrary 
exponent, as proposed by Antonov \cite{Antonov}, who only
applies perturbation theory. 
One could try to determine the arbitrary exponent from universality constraints,  
like in Ref.~\cite{I-EPL}. 
The stochastic approach deserves to be explored further.

After considering sophisticated methods of nonlinear dynamics, let us briefly comment on 
a simple method of obtaining the mass function of objects
formed by gravitational collapse, namely, 
the Press-Schechter method \cite{Sahni-C}. This method is based on the spherical collapse 
model, which is not realistic in three dimensions. In fact, Vergassola {\em el al} 
\cite{V-Frisch} find that the Press-Schechter mass function does not agree with 
the mass function derived from the adhesion model in more than one dimension. They  
define this mass function referring it to mass condensations in balls of fixed size, 
given that there are extended objects of arbitrary size, such as filaments, 
in more than one dimension. 
With a similar definition of the mass function, the form of this function has been studied 
in the distributions generated by cosmological $N$-body simulations \cite{I4,MN}. The 
result is a mass function of Press-Schechter type but with a fixed power-law 
exponent that is independent of $n$ (the initial power-spectrum exponent). 
This is contrary to 
the Press-Schechter approach.

One can see that 
the Press-Schechter is, in a way, analogous to the Peebles approach, in the 
sense that the latter also finds, 
under questionable assumptions, a simple relation, in this case, a relation between  
$n$ and another important exponent, namely, the $\xi$-exponent $\g$. However, 
it seems that simple approaches to the nonlinear gravitational 
dynamics that attempt to bypass its true complexity have limited scope.

\section{Discussion}


We have reviewed the main ideas and theories that have led to the current understanding 
of the geometry of the cosmic structure. The appealing denomination of it as a ``cosmic 
web'' is most appropriate in regard to its three-dimensional appearance, but some important properties, related to its fractal geometry,
are independent of the web morphology and already appear in 
one-dimensional cosmological models, in which the geometry is much simpler.

We have also considered halo models, very briefly.
Halo models are based on the statistical properties of discrete point distributions rather 
than on the geometrical analysis of 
continuous mass distributions. However, we have shown that the 
correlations between points cannot be too simple and must consist of a hierarchy that
allows us to 
naturally connect them with fractal models. Of course, this connection can 
actually take place only in the limit of an infinite number density of points, in which 
we obtain a continuous mass distribution.

The abstract study of the geometry of mass distributions has been essentially 
a mathematical subject. Since it is well developed, 
we propose to take advantage of this body of knowledge and we argue 
that the cosmic web structure must be studied as a strictly singular 
but continuous mass distribution. The geometry of these distributions is not simple even 
in one dimension, because they have non-isolated singularities. 
However, in the one-dimensional case, every mass distribution can be described 
in terms of the mass distribution {\em function}, which is just a monotonic function 
that gives the mass distribution by differentiation (it must be differentiable almost
everywhere). If the function is continuous, so is the mass distribution.
It is not easy to generalize this construction to higher dimensions, where various
geometrical and topological issues arise. At any rate, a study of the type of 
singularities in a continuous mass distribution is always possible. This is the 
goal of multifractal analysis, which is based on the analysis of the behavior of
{\em fractional} statistical moments $\M_q(l)$ when $l\ra 0$. 
In cosmology, only the integral moments $\M_n$ are normally considered, but they do 
not provide sufficient information.

The simplest strictly singular and continuous mass distribution 
consists of a uniform mass distribution on a fractal set, 
namely, on a self-similar set of the type of the Cantor set.
This mass distribution has just one kind of singularities. Therefore, it is a
monofractal, described by just one dimension, the Hausdorff dimension of the fractal set.
This type of fractal has a sequence of individual empty voids that is characterized by 
a particular form of the Zipf law.

Next in complexity, we consider a bifractal distribution; specifically, the type of 
distribution generated by the one-dimensional 
adhesion model. It is actually a non-continuous distribution  
with some interesting properties; namely, it does not contain 
any properly fractal set, with non-trivial Hausdorff dimension, but it is obviously
self similar. If we disregard its void structure, it just consists of a collection
of point-like masses, whose magnitudes follow a power-law, which is one
cause of its self-similarity. Another cause is the spatial distribution of the 
masses, which gives rise to a self-similar void structure.
In fact, this structure is crucial for the fractal nature, or rather bifractal, of 
this mass distribution, because the set of point-like masses, with $f(\a)=\a=0$, has 
nothing fractal to it, and the fractality is given by the other point of the bifractal 
spectrum, at $(1/h,0)$.

The transition from the preceding bifractal to a full multifractal is best perceived 
by focusing on the mass distribution function: let us substitute
the points of discontinuity, that is to say, the points 
with vertical segments in the graph, which correspond to $\a=0$, 
by points of non-differentiability, such that the derivative grows without limit.  
We so obtain a continuous monotone function 
that has weaker singularities, with $0<\a<1$.
This may seem an artificial procedure that is difficult to be realized in practice. 
On the contrary, any generic mass distribution is of this type. Moreover, 
these distributions normally have a non-trivial multifractal spectrum. 
If the graph of the function, besides not having vertical segments, neither has
horizontal segments, then there are no individual empty voids and 
the multifractal is non-lacunar. In this case, the voids 
look like the voids of the adhesion model distribution, but they have a 
more complex structure, 
because the multifractal spectrum contains a full interval with $\a>1$.

Naturally, the large-scale mass distribution is three-dimensional and, therefore, 
more complex, with notable morphological features. These features, in particular 
the shape of voids, inspired the early models of the cosmic web, e.g, 
the Voronoi foam model. 
Nowadays, important information on the structures of matter clusters and voids
can be obtained from the multifractal analysis of $N$-body 
simulations of the dynamics of cold dark matter alone or with baryonic matter, 
in combination with the multifractal analysis of galaxy catalogues.
The most obvious facts are, first, the self-similarity of the structures, and, 
second, that the multifractal spectra are incompatible with monofractal distributions.
We also notice the concordance of the multifractal spectra of cold dark matter and 
baryonic matter, to the extent allowed by the quality of the data. 
This concordance supports the 
hypothesis of a universal multifractal spectrum for the structure generated by 
gravitational clustering of both dark and baryonic matter.

Of course, this hypothesis challenges us to explain how such universality arises.
Some features of the multifractal spectrum can be easily related to
natural properties of the mass distribution. For example, $\mathrm{max} f = 3$ is related
to the non-lacunar structure or full support of the mass distribution. 
It can be explained by the adhesion model, which is a reliable approximation 
for the rough structure of voids (although not good enough for predicting the 
detailed multifractal spectrum for $\a > 3$).
On the opposite side of the multifractal spectrum, we have that $\a_{\mathrm{min}}=1$.
This value disagrees with the adhesion model prediction ($\a_{\mathrm{min}}=0$), and,
indeed, the adhesion model is a poor approximation for the formation of strong
mass concentrations. However, the basic condition of boundedness of the gravitational 
potential imposes $\a_{\mathrm{min}}=1$. In fact, a {\em monofractal} of dimension 
equal to one was 
proposed by Mandelbrot for a similar reason, namely, the moderate value of cosmic velocities 
\cite[\S 9]{Mandel}. But certainly the 
large-scale mass distribution is not monofractal, and the argument, properly 
considered, only sets a lower bound to the local fractal dimension.

To go beyond these two properties of the mass distribution, we need to understand better
the dynamics of structure formation in cosmology. It is useful to begin with 
a further analysis of the adhesion model, that is to say, 
with a study of Burgers turbulence, because the 
hydrodynamic type of nonlinearity of the dynamical equations proves that the 
methods of the theory of turbulence are appropriate. We consider 
the Richardson energy cascade, the Kolmogorov hypothesis, and the intermittency of 
the velocity field. 
The Burgers turbulence with scale invariant initial conditions lends itself to 
a fairly complete analysis and one can derive the 
form of the function $\z(q)$ that encodes the type of intermittency.

However, a treatment of the dynamics of structure formation that only involves 
the velocity field (because it is based on the Zeldovich approximation) 
is bound to be insufficient. A crucial problem of the adhesion model is that 
decaying Burgers turbulence does not conserve energy. 
This can be amended by studying forced Burgers turbulence, 
in the context of stochastic dynamics. The stochastic Burgers (or KPZ) equation gives rise 
to a scale-invariant asymptotic state that is independent of the initial conditions. 
The techniques based on dynamical scaling are powerful, but perturbation theory cannot deal 
with the three-dimensional problem and one has to apply non-perturbative methods. Simple
universality arguments lead to a constraint on $\z(q)$ and hence to a reasonable range
for the two-point correlation exponent $\g$.  But it is doubtful that these arguments 
can be pushed much further while keeping within the scope of Burgers turbulence.

Therefore, we must keep in mind the methods of turbulence but deal with 
the full gravitational dynamics. A lot of good work has been done on 
nonlinear gravitational dynamics and we cannot even mention most of it,
so we have limited ourselves to two relevant approaches: the classic Peebles closure 
approach and Gurevich-Zybin's theory of nondissipative gravitational singularities.
The full form of Peebles' closure approach involves some questionable hypotheses,
but we have focused precisely on the stable clustering hypothesis
for the highly nonlinear regime. This most natural hypothesis gives 
a prominent role to the conditional mass density and, hence, to fractal 
geometry. 
For the calculation of power-law exponents, namely, fractal dimensions, 
the stable clustering hypothesis falls short and additional assumptions are necessary.

Definite power-law exponents of gravitational singularities are provided by 
the Gurevich-Zybin theory, which consists in an appropriate treatment of multistreaming in
collisionless gravitational collapse.
However, this treatment can only obtain exponents for
singularities that form in isolation. We find no simple way to relate the exponents obtained 
by this theory to the dimensions in the multifractal spectrum of the real cosmic web. 
At any rate, the derivation 
of power-law singularities from the full dynamics of gravitational collapse 
is certainly an advance over the zero-size singularities of the adhesion model.

In conclusion, the problem of explaining the geometry of the cosmic web on a 
dynamical basis has only been solved partially. This problem seems to have a
similar status to the problem of the geometry of the flow in
fluid turbulence. Hopefully, the current developments 
in the study of nonlinear dynamics will bring further progress.

\conflictsofinterest{The author declares that there is no conflict of interest regarding the publication of this paper.}

\reftitle{References}

\end{document}